\documentstyle[aps,psfig,manuscript]{revtex}
\begin{document}
\title{Additive and Multiplicative Noise Driven Systems in 1 + 1 Dimensions:
Waiting Time Extraction of Nucleation Rates}

\author{Surujhdeo Seunarine\footnote{physsse@cantua.canterbury.ac.nz}}
\address{Department of Physics and Astronomy, The University of Canterbury, 
 \\ Private Bag 4800, Christchurch, New Zealand}
\author{Douglas W. McKay\footnote{mckay@kuhub.cc.ukans.edu}}
\address{Department of Physics and Astronomy, The University of Kansas, 
 \\Lawrence, KS 66045, U.S.A}

\maketitle

\begin{abstract}
We study the rate of true vacuum bubble nucleation
numerically for a $\phi^4$ field system coupled to a source of thermal noise. 
We compare in detail the cases of additive and multiplicative noise.  We pay 
special attention to the choice of initial field configuration, showing the
advantages of a version of the quenching technique.  We advocate a new method
of extracting the nucleation time scale that employs the full distribution of
nucleation times.  Large data samples are needed to study the initial state
configuration choice and to extract nucleation times to good precision.
The 1+1 dimensional models afford large statistics samples in reasonable 
running times.  We find that for both additive and multiplicative models, 
nucleation time
distributions are well fit by a waiting time, or gamma, distribution for all
parameters studied.  The nucleation rates are a factor three or more slower
for the multiplicative compared to the additive models with the same 
dimensionless parameter choices.  Both cases lead to high confidence level
linear fits of $ln\:{\tau}\:vs.\:T^{-1}$ plots, in agreement with semiclassical
nucleation rate predictions.
\end{abstract}

\section{Introduction}
In 1969 Langer \cite{langer} gave a first-principles formalism for calculating the decay rates 
of metastable states which was applicable to many types of nucleation and growth processes.
Several limiting cases were discussed, including those of a small energy difference between 
the two phases and overdamped and underdamped systems. 
Vacuum stability in scalar field theory was discussed by Lee and Wick \cite{lw}in 1974. 
In contemporaneous work, the cosmological consequences of the decay of metastable vacuum 
states were investigated by Zel'dovich, Kobzarev, and Okun \cite{zko} and was further 
developed by Voloshin, Kobzarev, 
and Okun \cite{vko}.  These authors investigated the idea of bubbles of a true vacuum
phase nucleating and expanding in a phase of the false vacuum. 
In 1981 Guth \cite{guth} proposed that sufficient supercooling of the false
vacuum before its decay could lead to enough exponential growth of the scale
factor (inflation) to solve the flatness and homogeneity problems.  This idea
has itself undergone significant evolution ever since.

A formalism for calculating nucleation rates, first using semiclassical
methods and then including 
quantum corrections, was worked out by Coleman \cite{coleman} and Coleman and 
Callan \cite{colemancallan} in 1977.  It was in these latter works that a 
Lorentz-invariant account of tunneling was 
given.  This technique uses a finite-action classical field configuration in 
Euclidean space, called the {\it bounce}.  The bounce
solution is the solution which minimizes the action.  They also
demonstrated how to treat zero modes in the decay rate calculation as well
as how to
determine the critical size of fluctuations which lead to bubbles that grow. 
In 1983 Linde \cite{linde1} generalized the study to finite temperatures.

Calculation of decay rates involve solving the equation of motion of the
scalar field. For the types of potentials which give rise to metastable
vacuum states, the equations of motion are non-linear in the field and, unless 
severe approximations are made, analytic solutions cannot be found. 
The detailed out-of equilibrium, time dependent aspects of metastable
vacuum decay can only be studied through numerical techniques. 
False vacuum decay at finite temperature has been studied numerically through
the use of Langevin equations \cite{valls}\cite{hume}\cite{algl}\cite{haas}. 
In these works the stochastic term in the Langevin equation was
in the form of additive noise, a field independent, white noise driving term.
Theoretical work \cite{morikawa}\cite{gleiserramos}\cite{GM} 
\cite{boyanovsky}, however, 
suggest that the noise term may be more complicated and in general may be
multiplicative and colored. We apply our study of Langevin dynamics to
the range of parameters appropriate to the semiclassical regime, where
there is reason to believe it makes sense\cite{hume}\cite{habib0}.  This 
way we can be sure about the conclusions drawn from our new
nucleation time method and about the comparison between the the additive 
and multiplicative noise models.

We study numerically the decay of the false vacuum in $1+1$ dimensions, 
focusing on the question of bubble nucleation rate.  We consider in detail 
the effects of two different types of noise terms: additive, where the 
nucleation rate has been studied before\cite{hume} in $1+1$  dimensions, and 
multiplicative, which has not been studied in numerical detail in this context.
Working in one spacial dimension affords several advantages: the system is
theoretically simple, with renormalization not an issue, for example
\cite{parisi}; the extraction of the nucleation time scale by the new method
we advocate requires generation of large data samples that can be achieved
in reasonable times in a $1+1$ dimensional system;  this simple system
allows us to explore relatively quickly the influence of initial state
configuration
choice on nucleation time distribution; the good statistics allows us to make
a sharp comparison between the results generated with our exploratory probe
of the multiplicative model and those generated with the additive model.

Given our objective we are immediately confronted by the question of the
proper definition of nucleation time.
In order to make full use of the nucleation time data generated with each
choice of input parameters (temperature, asymmetry parameter, driving 
amplitude and viscosity coefficient), we propose a new description of 
nucleation time based on a classic {\it waiting time} distribution
\footnote{An exponential decay can be fitted to the \emph{tails} of our
distributions, but this would be misleading, as we discuss in Sections
4 and 5 }. 
With this description of the distribution of nucleation times from the 
simulation data, we extract characteristic decay times for each parameter
set. We find that the relationship between nucleation time and temperature, 
derived from semiclassical calculations, are reproduced by both Langevin
models.  We also find that nucleation times are systematically a factor
of three or more longer in multiplicative noise models than in additive
models, given the same values of scaled input parameters and the same
initial state preparation in both models.

In the next section, we briefly review the ideas of vacuum transition in
zero and non-zero temperature cases.  In Section 3 we describe our numerical
methods, including the preparation of the initial state of the field.  In
Section 4 we describe our procedure for fitting our 
data and extracting nucleation rates and present our results.
In Section 5 we discuss our results and draw conclusions and comment on
open questions. Three appendices provide further details and supporting data
for material presented in the text.

\section{Quick Review of Nucleation Rates}
\begin{figure}[htbp]
\centerline{
        \psfig{file=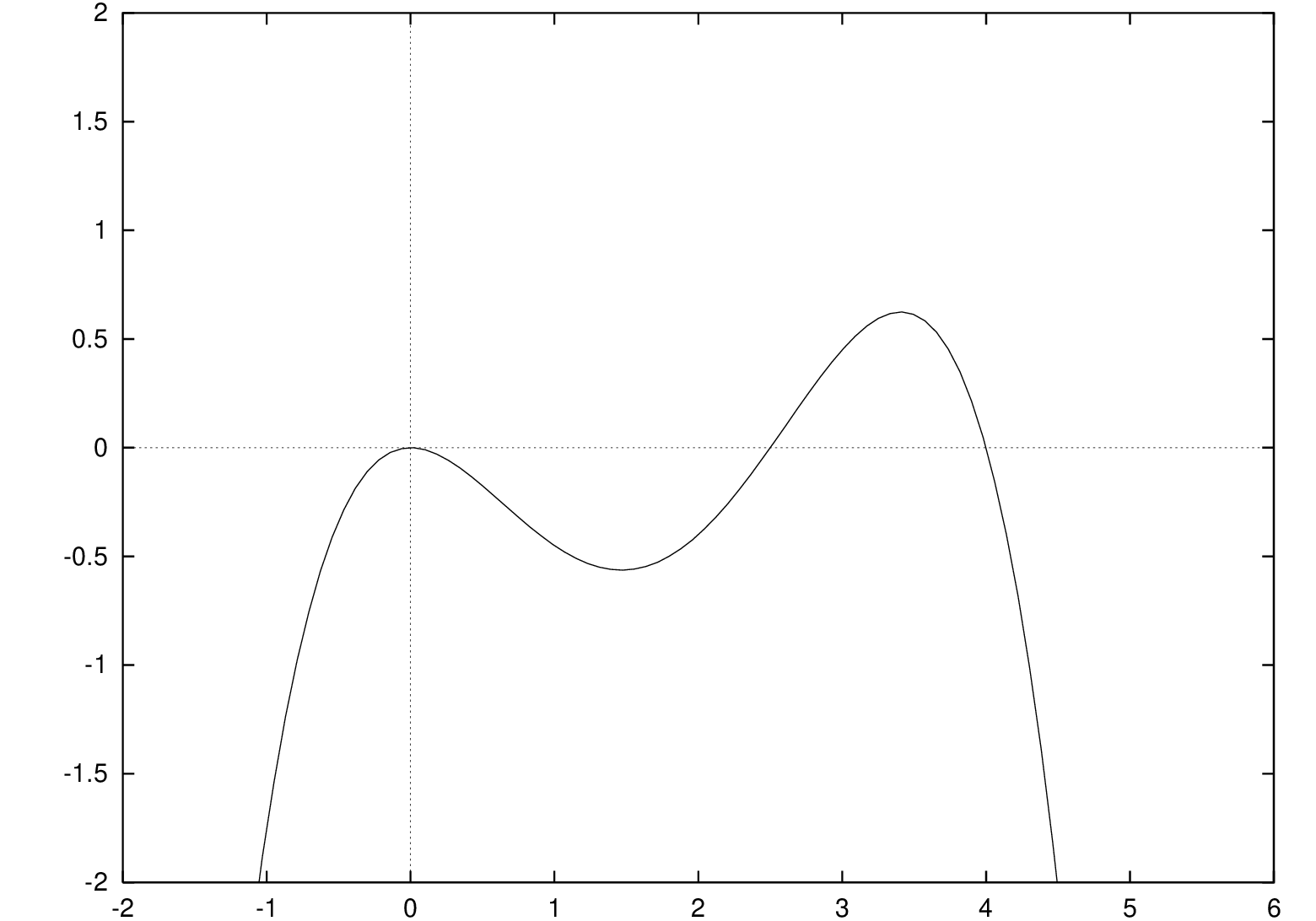,height=4in}}
        \caption{The inverted potential, $-V(\phi)\:vs\:\phi$. $\phi_1\approx 2.5$.}
        \label{figure1}
\end{figure}

\begin{figure}[h]
\centerline{    
        \psfig{file=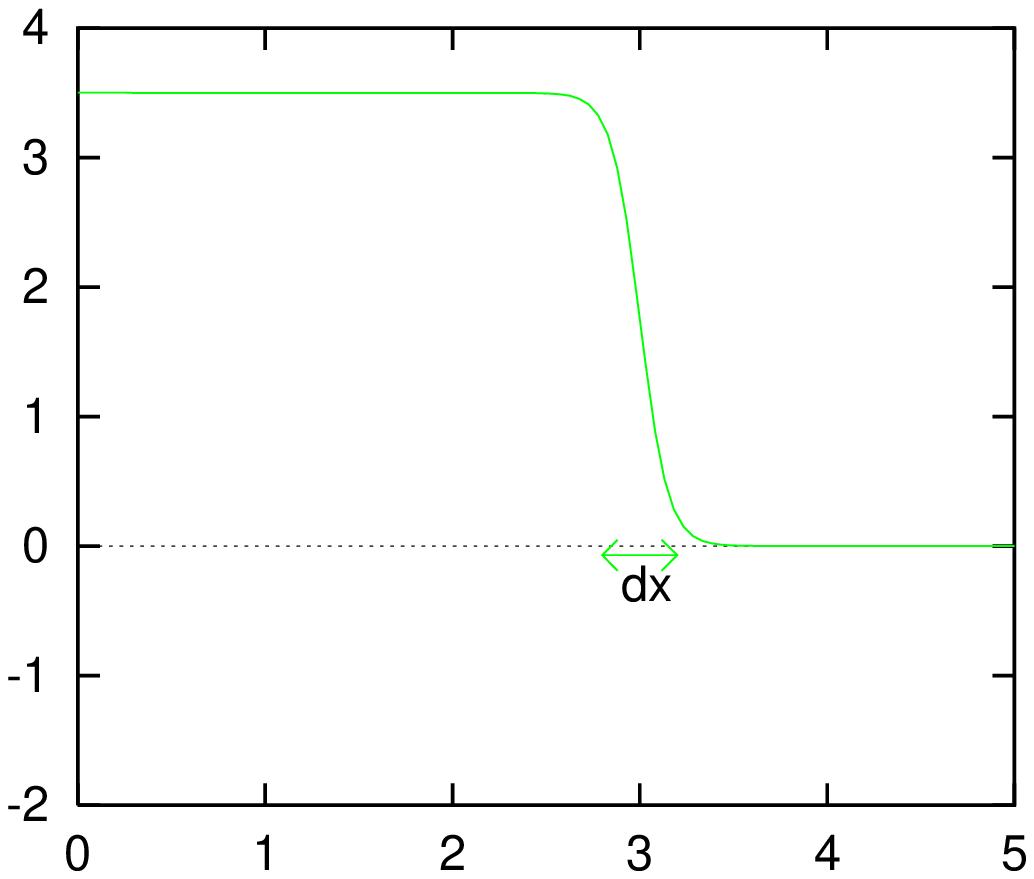,height=4in}}
        \caption{The bubble profile, $\phi(\rho)$, in the thin wall approximation. $dx$ is
        the narrow transition region or {\it wall}}
        \label{figure2}
\end{figure}

Consider a self-interacting scalar field theory and suppose the field
starts off with a value $\phi=0$, which corresponds to the
metastable minimum of the  effective potential (hereafter referred
to as the potential).  Classically, transition to the global minimum at 
$\phi=\phi_0$ is forbidden,
but there is a 
non-zero probability for quantum tunneling through the barrier with bubbles
of the true vacuum 
appearing in the metastable phase. 
 The decay probability per unit volume can be expressed by the functional
integral\cite{coleman}\cite{colemancallan}
%%%%%%%%%%%%%%%%%%%%%%%%%%%%%%
\begin{equation}
     \Gamma=\int D\phi e^{-S_E[\Phi]},
\label{path_int}
\end{equation}
%%%%%%%%%%%%%%%%%%%%%%%%%%%%%%
where $S_E$ is the Euclidean action,
%%%%%%%%%%%%%%%%%%%%%%%%%%%%%%
\begin{equation}
     S_E=\int d\tau d^3x[\frac{1}{2}(\frac{\partial{\phi}}{\partial{\tau}})^2+
\frac{1}{2}(\frac{\partial{\phi}}{\partial{x}})^2+V(\phi)].
\label{euclidean_action}
\end{equation}
%%%%%%%%%%%%%%%%%%%%%%%%%%%%%%
The equation of motion is 
%%%%%%%%%%%%%%%%%%%%%%%%%%%%%
\begin{equation}
     \frac{\partial^2{\phi}}{\partial{\tau^2}}+\frac{\partial^2{\phi}}{\partial{x^2}}=
        \frac{dV}{d\phi}.
\label{eqn_motion}
\end{equation}
%%%%%%%%%%%%%%%%%%%%%%%%%%%%%%
Notice the positive sign in front of the term on the right.  This is the
equation of motion 
of the field in the potential $-V(\phi)$, exemplified in Fig. \ref{figure1}. 
Now the boundary conditions have to be suitably chosen.  The field starts off
at $\phi=0$ in the 
inverted potential and gets to the escape point $\phi_1$.  In the case of
degenerate minima
the path integral is dominated by the non-trivial minimum energy solution of 
the equation 
of motion, the instanton\cite{polyakov}\cite{rajaraman}. 
For non-degenerate minima the 
non-tivial solution to the equation of motion in Euclidean space corresponds
to the field starting at 
the unstable minimum, $\phi=0$, at $\tau=-\infty$, 'bouncing' off the escape point at
$\tau=0$ and asymptotically approaching $\phi=0$ again, as $\tau=+\infty$, the {\it bounce} solution.  
%%%%%%%%%%%%%%%%%%%%%%%%%%%%%%
Semi-classically, the path integral in Eqn. \ref{path_int} gives the
tunneling rate per unit volume as,
%%%%%%%%%%%%%%%%%%%%%%%%%%%
\begin{equation}
        \Gamma=Ae^{-S_B},
\label{rate}
\end{equation}
%%%%%%%%%%%%%%%%%%%%%%%%%%%%%%
where $S_B$ is the action, Eqn. \ref{euclidean_action}, evaluated with the solution of the equation of
motion with boundary conditions $\phi(\tau=-\infty,x)=\phi(\tau=+\infty,x)=0$.  The 
coefficient A contains quantum corrections to the tunneling rate.
In most cases it is not possible to solve the equation of motion  analytically
with the appropriate 
boundary conditions. However there are limiting cases \cite{langer}\cite{coleman} where exact
analytic solutions can be found. It was shown \cite{colemanglasermartin} that the
spherically symmetric, $O(4)$, invariant bounce is the one with the least 
action.
Then with the
boundary conditions for the $O(4)$ symmetric bounce the general solution for $\phi(\rho)$ is of the so-called {\it kink} 
form

%%%%%%%%%%%%%%%%%%%%%%%
\begin{equation}
        \phi=\frac{1}{2}\phi_o[1-tanh(\frac{\rho-R}{dx})],
        \label{twa_solution}
\end{equation}
%%%%%%%%%%%%%%%%%%%%%%%
where $\rho=(\tau^2+{\bf x}^2)^\frac{1}{2}$ and $dx$ is the thickness of
the transition region.  This solution, valid when the difference between the
energy density in the true vacuum and that in the false vacuum is small\cite{coleman}, is called
the thin wall solution and the general form is depicted in
Fig. \ref{figure2}. 
As the name of the approximation suggests, there is a sharp transition region
(or thin wall) between
the true and false vacuum phases. The interpretation of Fig. 
\ref{figure2} is that somewhere in space, 
at $\rho=0$, a bubble of the true vacuum, $\phi=\phi_0$, is formed. Away from
the center
of the bubble there is a transition region, which is very narrow on the thin
wall approximation.
Beyond this transition region, and in particular as $\rho\rightarrow\infty$,
the field is in the false vacuum state $\phi=0$.  The parameter $R$ appearing in 
Eqn. \ref{twa_solution} is the radius of a critical 
size bubble which can be easily calculated in the thin wall
approximation.  

Tunneling at finite temperature can be studied using all of the above 
formalism if one
uses the imaginary time technique of Matsubara \cite{matsubara}, 
exploiting the equivalence between
four dimensional field theory at finite temperature and Euclidean field theory
in four dimensions with periodicity or anti-periodicity in one of the
dimensions. The period
is $\beta$ where $\beta=1/T$. One therefore has the requirement that 
$\phi(\tau,\rho)=\phi(\tau+\beta,\rho)$. For sufficiently high
temperatures this additional condition gives the
finite temperature action 
%%%%%%%%%%%%%%%%%%%%%%%%%%%%%%%%%%%%%
\begin{equation}
        S_E=\frac{S_3}{T},
\label{ft_action}
\end{equation}
%%%%%%%%%%%%%%%%%%%%%%%%%%%%%%%%%%%%
where $S_3$ is the three dimensional Euclidean action and $T$ is the
temperature.  Eqn. \ref{ft_action} is valid at sufficiently high
temperatures, higher than the temperature at which  
the period, $\beta$, of the bounce solution is equal the critical
radius, $R$, of a bubble\cite{linde1}.
%%%%%%%%%%%%%%%%%%%%%%%%%%%%%%%%%%%%%%%%%
The solution to the equation of motion with the least action is now $O(3)$
symmetric and the equation of motion can be solved in the thin wall 
approximation as above.
\section{Numerical Studies and Time Evolution}
The theory of false vacuum decay outlined in the preceding section made clear 
that nucleation can be studied analytically only in a few limiting
cases. Out of equilibrium time evolution properties can only be studied
in detail using numerical techniques.  
Nucleation has been studied numerically through the use of Langevin equations. 
Early calculations, for example\cite{hume}, 
used a Langevin equation for the time evolution of the field based solely on
phenomenological grounds. Other  
studies have attempted to derive Langevin type equations from quantum field
theory in the semiclassical limit. Typically the high frequency modes are
treated as the thermal bath for the low frequency modes, which is identified as
``the system''\cite{gleiserramos}\cite{GM}\cite{KMT}.
  
For a scalar field theory in one
space and one time dimension, a phenomenological Langevin type equation can 
be written as,  
%%%%%%%%%%%%%%%%%%%%%%%%%%%%%%%%%%%%%%%%%%%%%%%%%%%%%%%%%%%%
\begin{equation}
\frac{\partial^2\phi(x,t)}{\partial t^2}-\frac{\partial^2\phi(x,t)}{\partial x^2}+
        \eta\frac{\partial\phi(x,t)}{\partial t}=-V'(\phi)+\xi(x,t,T)
        \label{full_add_noise1}
\end{equation}
%%%%%%%%%%%%%%%%%%%%%%%%%%%%%%%%%%%%%%%%%%%%%%%%%%%%%%%%%%%%
where $\xi(x,t)$ is a random force term originating from the coupling of the field to a thermal bath.
The thermal noise term, $\xi(x,t)$, and the viscosity, $\eta$, are related through the fluctuation
dissipation theorem\cite{kubo}\cite{chandrasekhar}\cite{parisi},
%%%%%%%%%%%%%%%%%%%%%%%%%%%%%%%%%%%%%%%%%%%%%%%%%%%%%%%%%%%%%%%%%%%%%%%%%%%
\begin{equation}
        <\xi(x,t)\xi(x',t')>=2T\eta\delta(x-x')\delta(t-t').
        \label{f_d}
\end{equation}
%%%%%%%%%%%%%%%%%%%%%%%%%%%%%%%%%%%%%%%%%%%%%%%%%%%%%%%%%%%%
Since Eqn. \ref{f_d} is central to our work and is used, but seldom discussed,
in the field theory literature, we provide details on its implementation 
and  its interpretation in our study in {\it Appendix A}.
%%%%%%%%%%%%%%%%%%%%%%%%%%%%%%%%%%%%%%%%%%%%%%%%%%%%%%%%%%%%
The appearance of the viscosity, $\eta$, in the equations of motion and its 
relation to the thermal fluctuations through Eqn. \ref{f_d}
can be understood by considering the example of Brownian motion.  
A Brownian particle suspended in a fluid is
subjected to random, microscopic, short-duration forces which, over a longer
period of time, 
change the macroscopic velocity of the particle.  At the same time the random
forces
in the direction opposite to the velocity of the particle serve to retard its
motion. The very   
forces, therefore, that are responsible for the macroscopic motion of the 
suspended particle also are the ones that inhibit the motion, namely 
give rise to the viscosity of the medium. In order to include both 
effects of the random forces, two independent terms
are introduced in the equation of motion; they are connected through the 
fluctuation-dissipation relation.

In Eqn. \ref{full_add_noise1} the fluctuation term is linear.  This is
referred to as
additive noise.  However there is no reason why one should not consider a
general case where
the noise is non-linear.  For example, a Langevin equation with a non-linear 
thermal fluctuation term has been derived, in a set of approximations
\cite{gleiserramos}, by considering a self interacting quantum scalar field.  
It was shown that, at finite temperature, if the short wavelength modes of
the field are separated 
out and the two-loop effective potential is used, then the effective action 
contains imaginary terms which are interpreted as coming from Gaussian 
integrations over auxiliary 
fields, $\tilde \xi$. These auxiliary fields appear in the the equation of
motion in the form of a random fluctuation term.  
We adopt a 
phenomenological version of this result in a $1+1$ dimensional multiplicative
noise model to
discover what features of the nucleation rate may differ from those of
the additive noise case.  
%%%%%%%%%%%%%%%%%%%%%%%%%%%%%%%%%%%%%%%%%%%%%%%%%%%%%%%%%%%%%%%%%%%%%%%%%%%%%%
\begin{equation}
\frac{\partial^2\phi(x,t)}{\partial
 t^2}-\frac{\partial^2\phi(x,t)}{\partial x^2}+
        \eta\phi^2(x,t)\frac{\partial\phi(x,t)}{\partial t}=-V'(\phi)+\phi(x,t)\tilde\xi(x,t,T).
        \label{full_mul_noise1}
\end{equation}
%%%%%%%%%%%%%%%%%%%%%%%%%%%%%%%%%%%%%%%%%%%%%%%%%%%%%%%%%%%%%%%%%%%%%%%%%%%% 
Notice that there is now a non-linear viscosity coefficient. In the high
 temperature limit the fluctuation-dissipation relation, Eqn. \ref{f_d}, 
is valid\cite{gleiserramos}.

We use the Langevin equations with linear and non-linear noise terms, and with
the usual fluctuation-dissipation theorem to compare the effects of these
types of noise on nucleation. This is the first such numerical study,
as far as we know.

We re-scale the equations so that all parameters appearing in them are 
dimensionless. We give the detail of the re-scaling in appendix B. 
We propagate the solution of the equation of motion using a staggered 
leap-frog method\cite{press}.  We have checked 
that the results do not depend on this method.  We also repeated simulation
runs on 
a lattice twice as large as the one reported here to check that the results
show no {\it finite volume} (or in 1-dimension finite length) effects. 
We use a lattice of size $L=100$ with a space step-size of 
$\delta x=.5$ and a time step of $\delta t=.1$ and with periodic boundary
conditions.  The
value of the field at the $n+1$ time step in terms of the values at $n$ and
$n-1$, and at position $m$ on the lattice is given, for the additive model,
Eqn.  \ref{full_add_noise1}, by
%%%%%%%%%%%%%%%%%%%%%%%%%%%%%%%%%%%%%%%%%%%%%%%%%%%%%%%%%%%%%%%%%%%%%%%%%%%%%%%%%%%%%%%%%%
\begin{equation}
        \phi_{n+1,m}=2\phi_{n,m}-\phi_{n-1,m}+\delta t^2(\nabla^2\phi_{n,m}-\eta\Pi_{n,m}
-V'(\phi_{n,m})+\xi_{m,n}),
        \label{discrete_eqn_motion}
\end{equation}
%%%%%%%%%%%%%%%%%%%%%%%%%%%%%%%%%%%%%%%%%%%%%%%%%%%%%%%%%%%%%%%%%%%%%%%%%%%%%%%%%%%%%%%%%
where,
%%%%%%%%%%%%%%%%%%%%%%%%%%%%%%%%%%%%%%%%%%%%%%%%%%%%%%%%%%%%%%%%%%%%%%%%%%%%%%%%%%%%%%%%%%
\begin{equation}
        \Pi_{n,m}=(\phi_{n+\frac{1}{2},m}-\phi_{n-\frac{1}{2},m})/\delta t,
\end{equation}
%%%%%%%%%%%%%%%%%%%%%%%%%%%%%%%%%%%%%%%%%%%%%%%%%%%%%%%%%%%%%%%%%%%%%%%%%%%%%%%%%%%%%%%%%%
and
%%%%%%%%%%%%%%%%%%%%%%%%%%%%%%%%%%%%%%%%%%%%%%%%%%%%%%%%%%%%%%%%%%%%%%%%%%%%%%%%%%%%%%%%%%
\begin{equation}
        \nabla^2\phi_{n,m}=(\phi_{n,m+1}-2\phi_{n,m}+\phi_{n,m-1})/\delta x
\end{equation}
%%%%%%%%%%%%%%%%%%%%%%%%%%%%%%%%%%%%%%%%%%%%%%%%%%%%%%%%%%%%%%%%%%%%%%%%%%%%%%%%%%%%%%%%%%  
Gaussian white noise is used for both the linear(additive) and non-linear(multiplicative) 
couplings to the thermal bath. We relate the noise and viscosity through Equation \ref{f_d}.
%%%%%%%%%%%%%%%%%%%%%%%%%%%%%%%%%%%%%%%%%%%%%%%%%%%%%%%%%%%%%%%%%%%%%%%%%%%%%%%%%%%%%%%%%%
The discrete form of this equation is
%%%%%%%%%%%%%%%%%%%%%%%%%%%%%%%%%%%%%%%%%%%%%%%%%%%%%%%%%%%%%%%%%%%%%%%%%%%%%%%%%%%%%%%%%% 
\begin{equation}
        <\xi_{m,n}\xi_{m',n'}>=\frac{2T\eta}{\delta x\delta t}\delta_{mm'}\delta_{nn'}
        \label{f_d2}
\end{equation}
%%%%%%%%%%%%%%%%%%%%%%%%%%%%%%%%%%%%%%%%%%%%%%%%%%%%%%%%%%%%%%%%%%%%%%%%%%%%%%%%%%%%%%%%%%
where the $\delta x$ and $\delta t$ terms are introduced to compensate for the lack of 
dimensionality of the Kroneker deltas.  $\xi$ is given by
%%%%%%%%%%%%%%%%%%%%%%%%%%%%%%%%%%%%%%%%%%%%%%%%%%%%%%%%%%%%%%%%%%%%%%%%%%%%%%%%%%%%%%%%%%
\begin{equation}
        \xi_{m,n}=\sqrt{\frac{2T\eta}{\delta x\delta t}}G_{m,n}
\end{equation}
%%%%%%%%%%%%%%%%%%%%%%%%%%%%%%%%%%%%%%%%%%%%%%%%%%%%%%%%%%%%%%%%%%%%%%%%%%%%%%%%%%%%%%%%%%
where $G_{m,n}$ is a Gaussian random number generator of width one.  The delta correlation
is implemented by using a different random {\it kick} at each time and at each point
on the lattice.  By looking at approximately 100 bubble
profiles we found that fluctuations which take the field at five contiguous lattice sites to 
the true vacuum value always grew to fill the entire lattice.  Therefore we chose this criterion 
to determine when a bubble was formed.
%%%%%%%%%%%%%%%%%%%%%%%%%%%%%%%%%%%%%%%%%%%%%%%%%%%%%%%%%%%%
\subsection{Initial Conditions}
%%%%%%%%%%%%%%%%%%%%%%%%%%%%%%%%%%%%%%%%%%%%%%%%%%%%%%%%%%%%
The initial conditions in such a numerical study have to be carefully chosen.
We looked at nucleation times with  random, uncorrelated initial field values
distributed on the lattice.  We also looked at nucleation times with short
distance correlated and Gaussian distributed initial field values.  We found
that with the first case there was a long delay time during which no bubbles 
were formed.  After this delay time there was a distribution of nucleation
times rather sharply peaked around some value.  With the second set of
initial conditions, i.e. correlated and Gaussian distributed, we found that
this delay time was sharply reduced, the distribution rising quickly and
peaking much earlier.  This is illustrated for samples of 5,000 bubbles
in Fig. \ref{figure5}.

\begin{figure}[hbtp]
\centerline{    
        \psfig{file=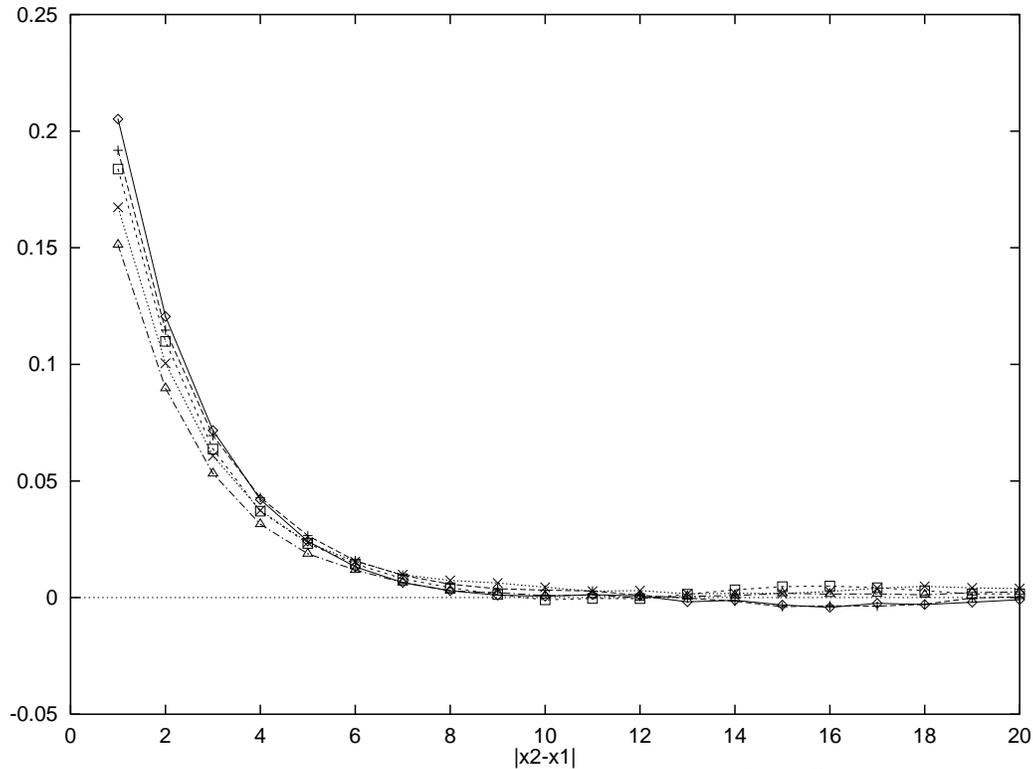,height=4in}}
        \caption{Field correlation $<\phi(x1)\phi(x2)>$ just before quenching for different temperatures 
        with $\diamond$'s representing the highest temperature and $\triangle$'s the lowest.}
        \label{figure3}
\end{figure}

\begin{figure}[hbtp]
  \vspace{9pt}
  \centerline{\hbox{ \hspace{0.0in} 
        \psfig{file=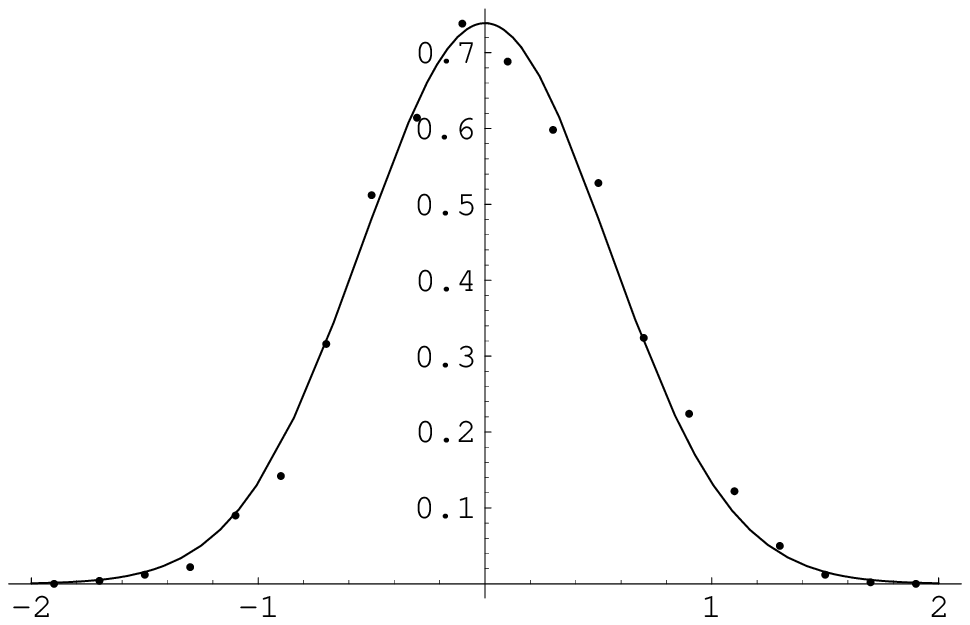,height=1.5in}
    \hspace{0.2in}
         \psfig{file=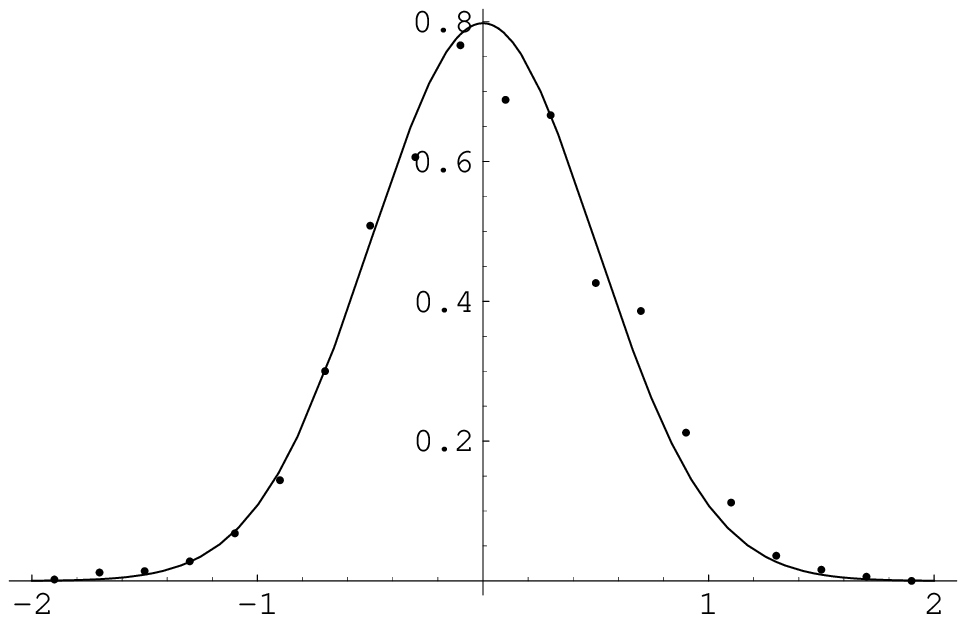,height=1.5in}
    }
  }
\vspace{9pt}
\hbox{\hspace{1.35in}{a} \hspace{2.10in} (b)} 
\caption{Gaussian distribution of initial conditions obtained using the
 quenching technique, $\frac{dN(\phi)}{d\phi}$ vs $\phi$,
for (a) additive noise and (b) multiplicative noise.  The points are data and
the curves are fits to Gaussians. Here the area under the curve between $\phi$
 and $\phi+\delta\phi$
 is the fraction of lattice points having the field value in that range.}
\label{figure4}
\end{figure}

\begin{figure}[htbp]
\centerline{    
        \psfig{file=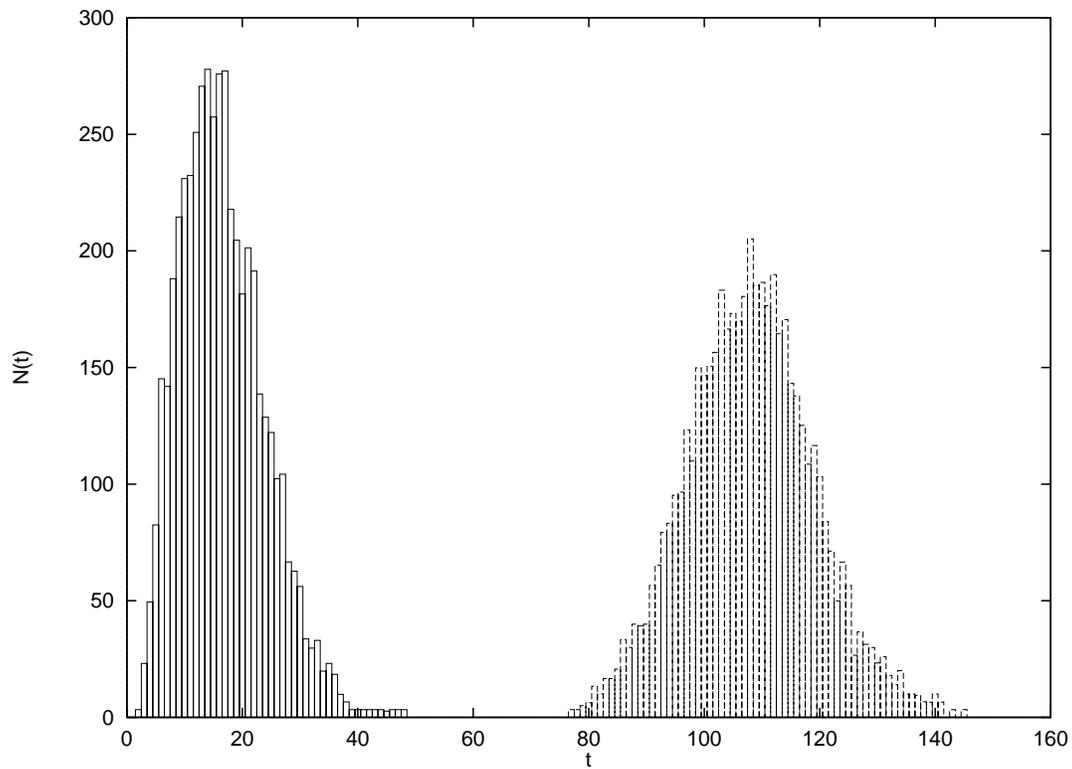,height=4in}}
        \caption{Distribution of nucleation times with quenching, left, and with
 random initial conditions, right}
        \label{figure5}
\end{figure}

The initial field values, whose correlations and distributions are
described by Figs. \ref{figure3} and \ref{figure4}
 are obtained using a {\it quenching} technique\cite{bettencourt}. The effect
of this initial state preparation on the bubble nucleation times is shown by the data indicated by the histogram
on the left  in Fig. \ref{figure5}.  We scatter random field values
on the lattice and propagate the solution of the equation of motion in a potential with only one minimum.  We
ensure that the curvature at the bottom of the potential closely matches the curvature of the
asymmetric double well of interest. After some time the field acquires the thermal distribution
with the desired short distance correlation.  At this point, which is $t=0$ in the simulation, 
the potential is quenched, i.e., the asymmetric potential is switched on.  Nucleation times are
recorded with reference to this time. Fig. \ref{figure4} shows the distribution of initial conditions just 
before quenching for both additive noise and multiplicative noise.

Summarizing, we find that when the initial state is not prepared by the 
quenching method just described, and random, uncorrelated, initial field
values are placed on the lattice, there is a long delay time for the first
bubble to appear, after which a peaked distribution of times develops.
The delay time apparently corresponds to the time needed for the field
to become correlated.  
Preparing the initial state so that it is correlated as shown in Figs. 
\ref{figure3} and \ref{figure4}, produces the change in the 
distributions illustrated in Fig. \ref{figure5}.  We use correlated
inital states prepared by the quenching technique in all of the data shown
and discussed in the following sections of the paper.

\subsection{Time Evolution}

\begin{figure}[t]
  \vspace{9pt}

  \centerline{\hbox{ \hspace{0.0in} 
    \psfig{file=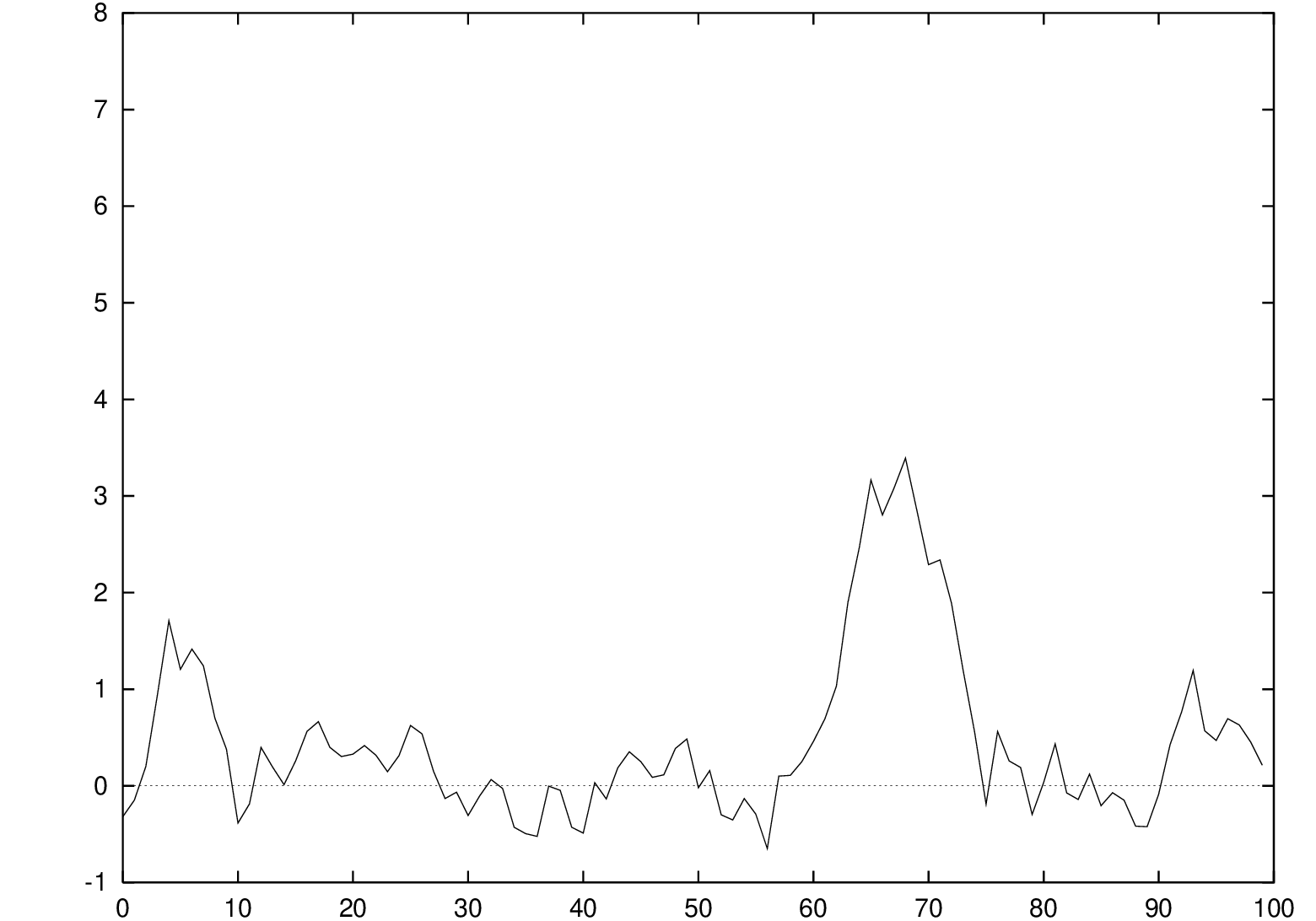,height=1.5in}
    \hspace{0.2in}
    \psfig{file=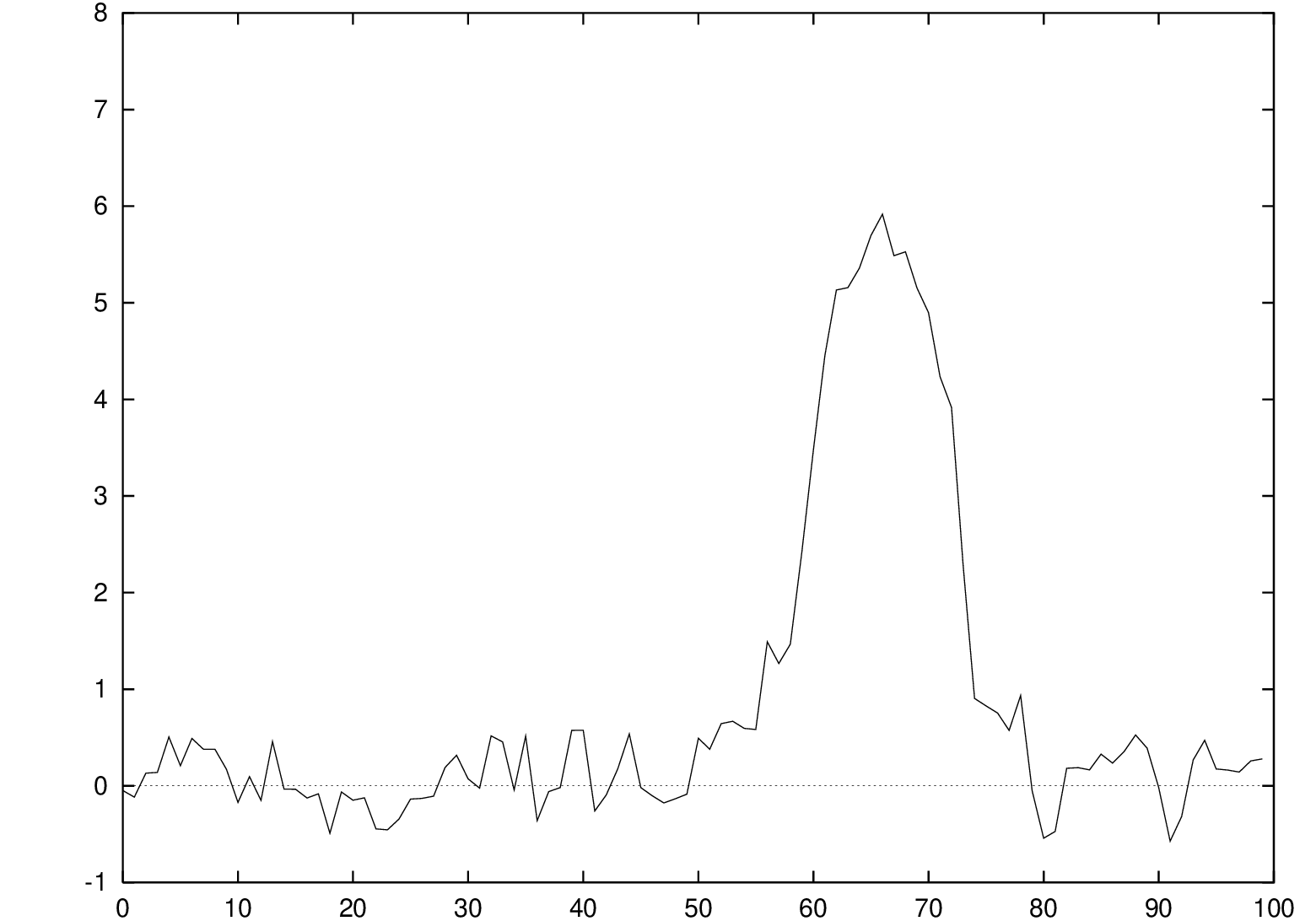,height=1.5in}
    }
  }
\vspace{9pt}
\hbox{\hspace{1.5in} ($t=100$) \hspace{1.75in} ($t=150$)} 
  \vspace{9pt}

  \centerline{\hbox{ \hspace{0.0in} 
    \psfig{file=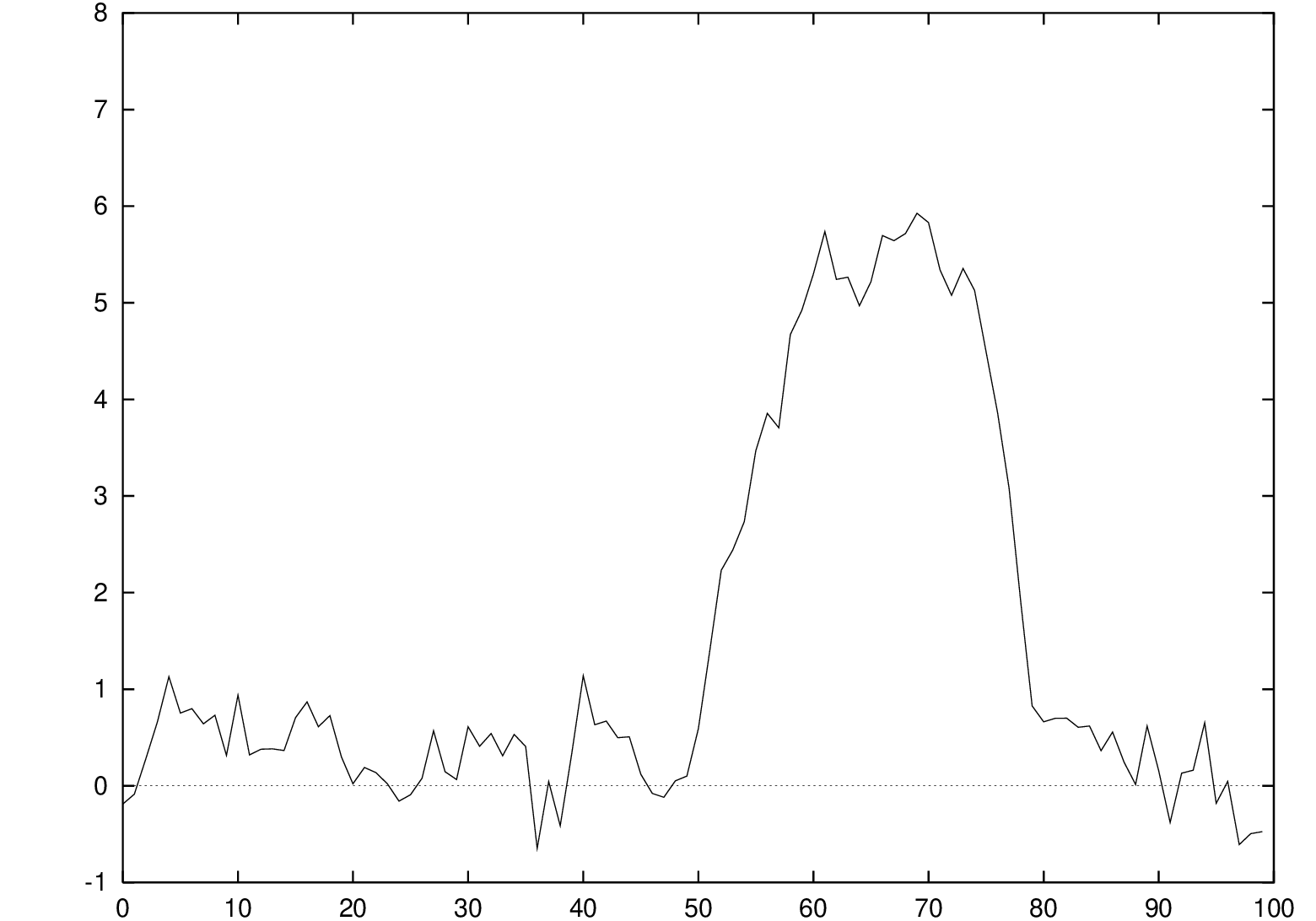,height=1.5in}
    \hspace{0.2in}
    \psfig{file=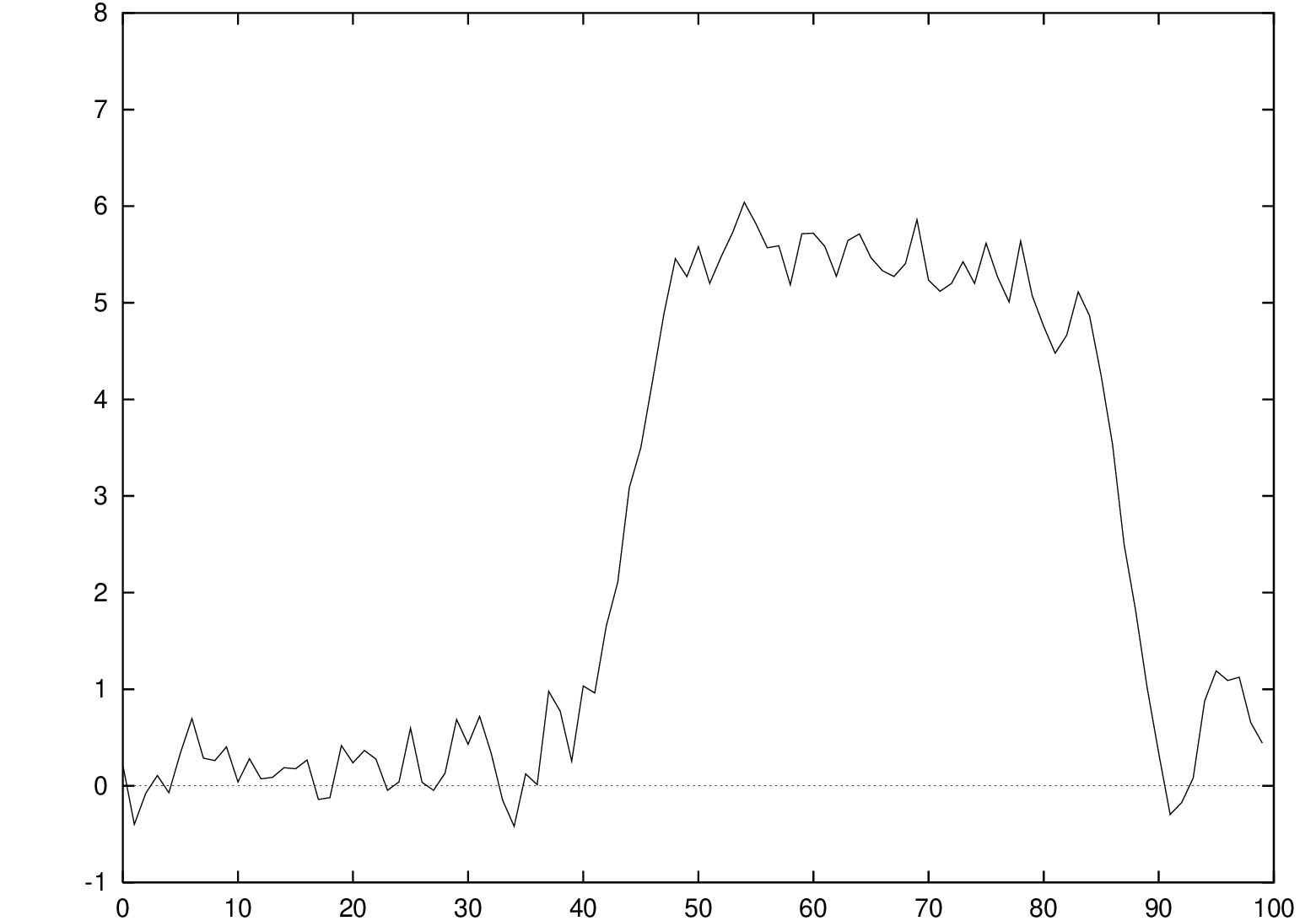,height=1.5in}
    }
  }
\vspace{9pt}
\hbox{\hspace{1.5in} ($t=200$) \hspace{1.75in} ($t=300$)} 
\caption{A series of snapshots of the lattice, $\phi(x)$ vs $x$, showing 
the formation of a bubble and its subsequent growth for an
additive noise system.  The absissa is the spatial
position and the ordinate is the scaled field value.}
\label{figure6}
\end{figure}

\begin{figure}[t]
  \vspace{9pt}

  \centerline{\hbox{ \hspace{0.0in} 
    \psfig{file=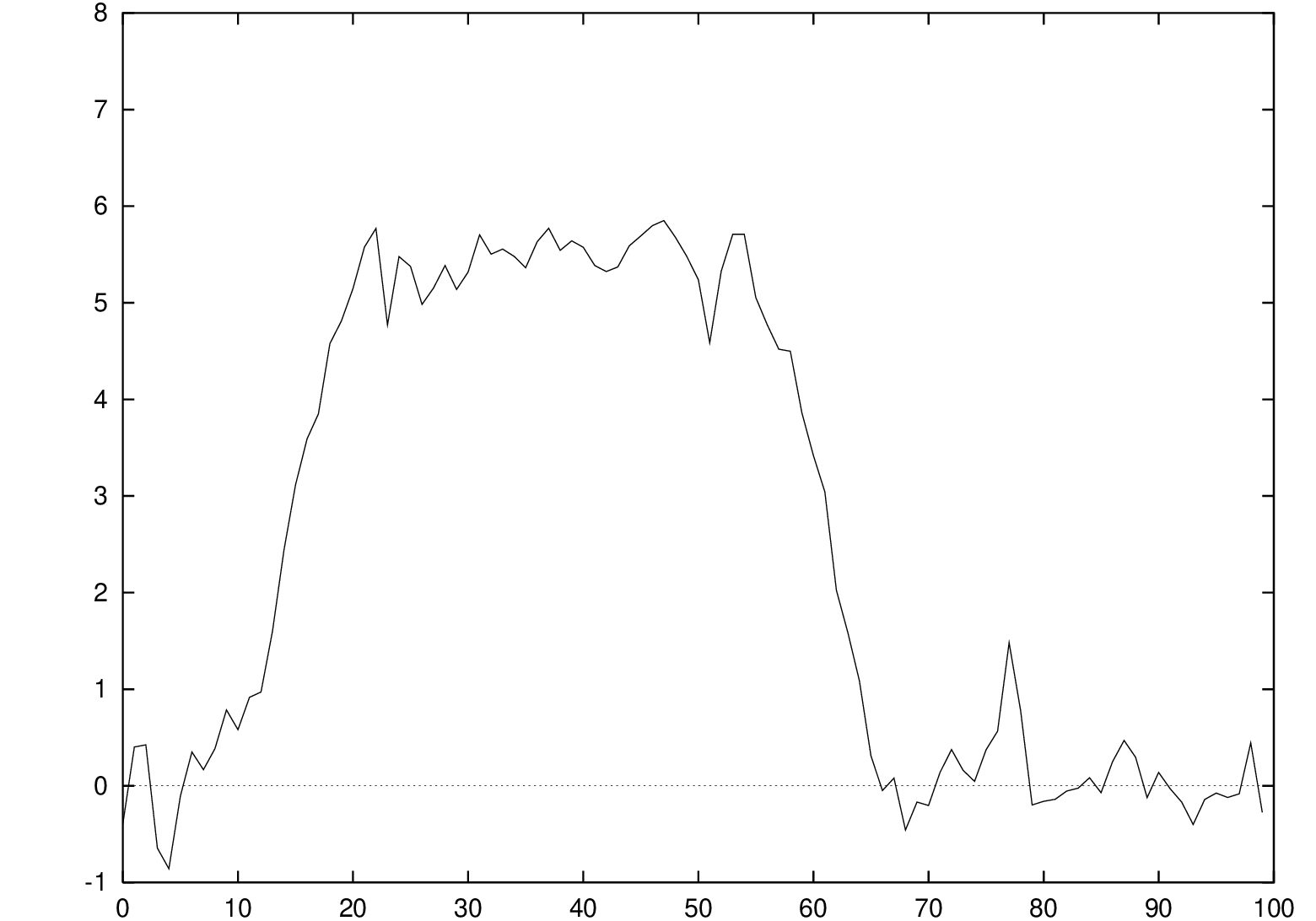,height=1.5in}
    \hspace{0.2in}
    \psfig{file=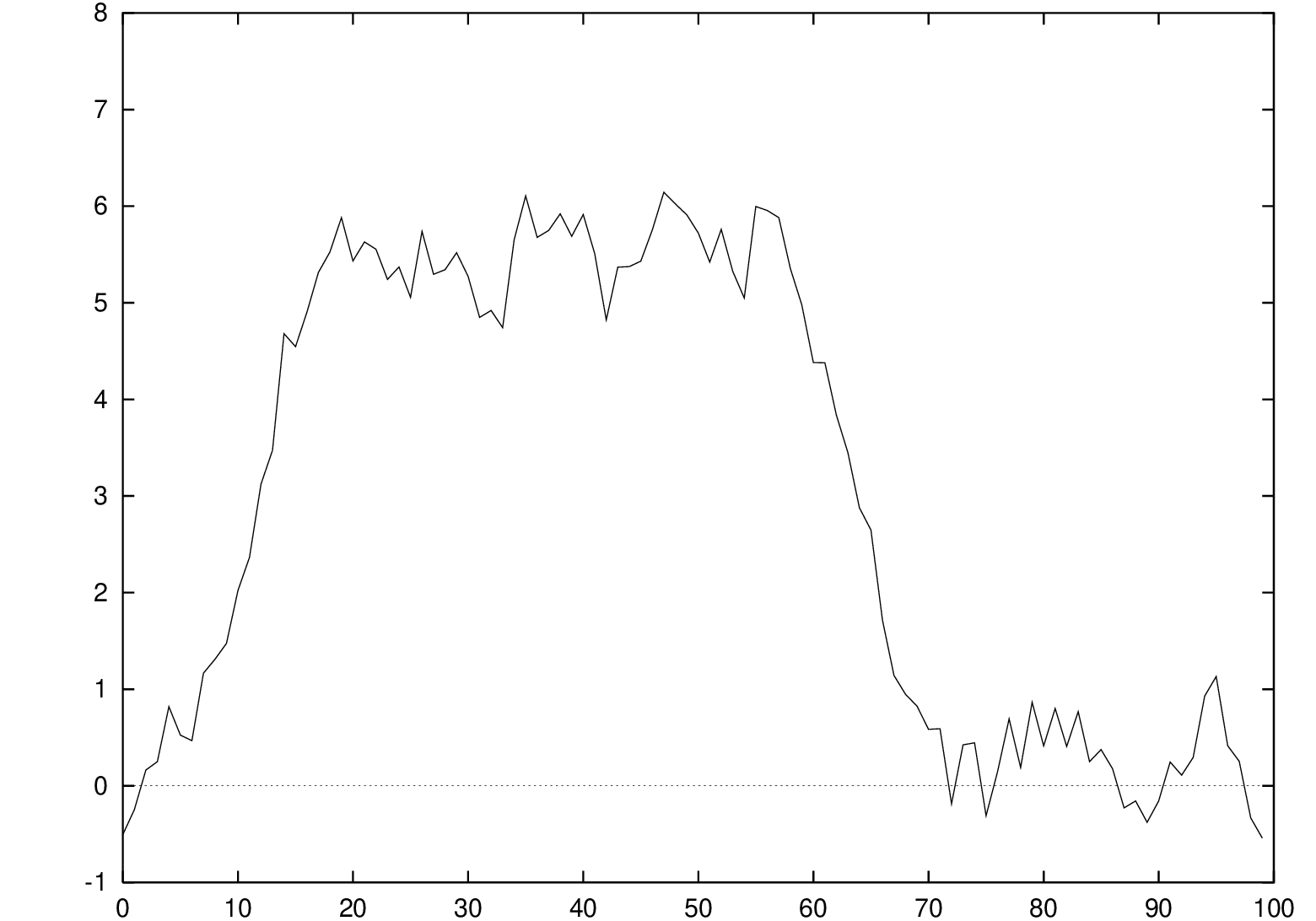,height=1.5in}
    }
  }
\vspace{9pt}
\hbox{\hspace{1.5in} ($t=901$) \hspace{1.75in} ($t=1100$)} 
  \vspace{9pt}

  \centerline{\hbox{ \hspace{0.0in} 
    \psfig{file=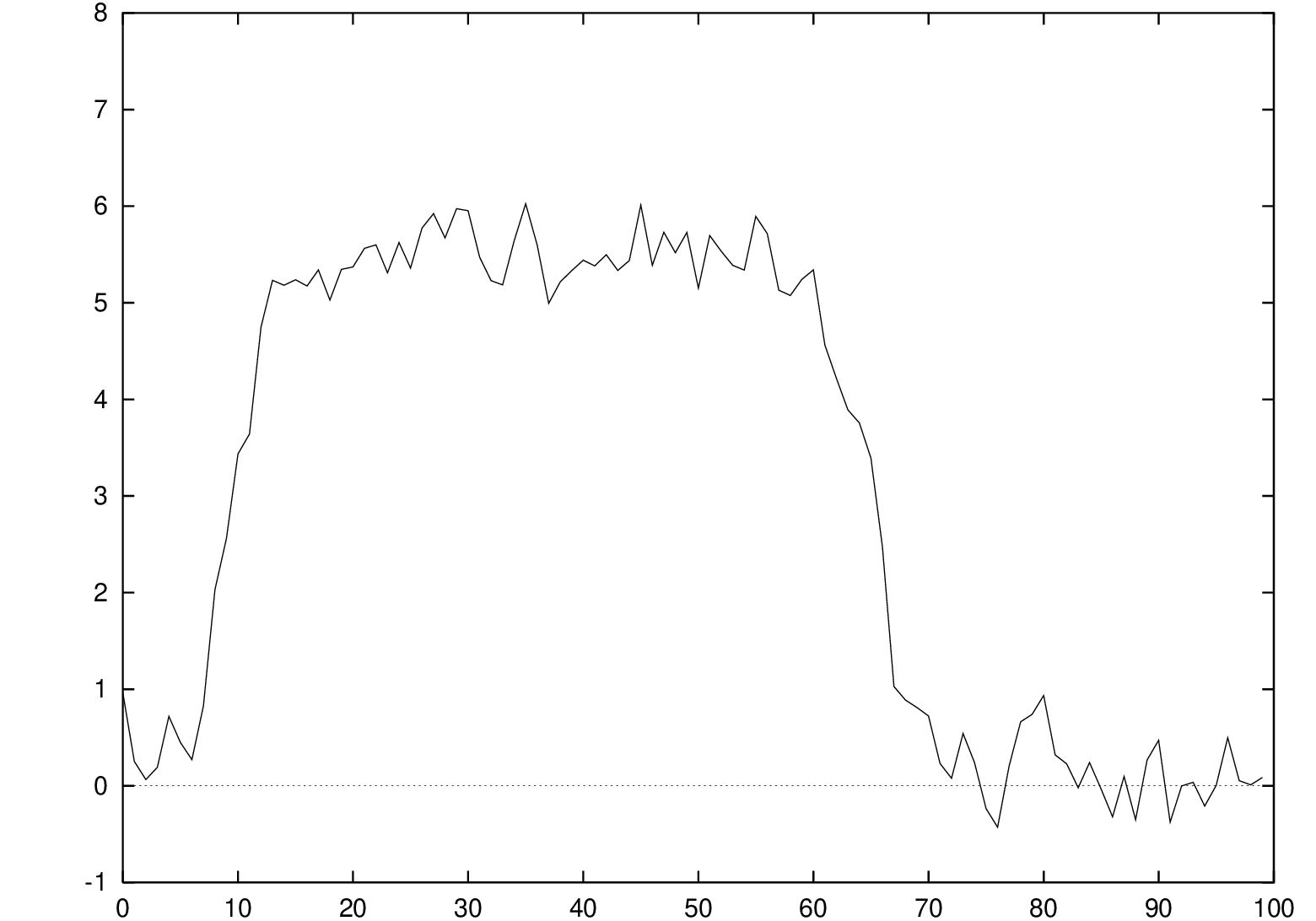,height=1.5in}
    \hspace{0.2in}
    \psfig{file=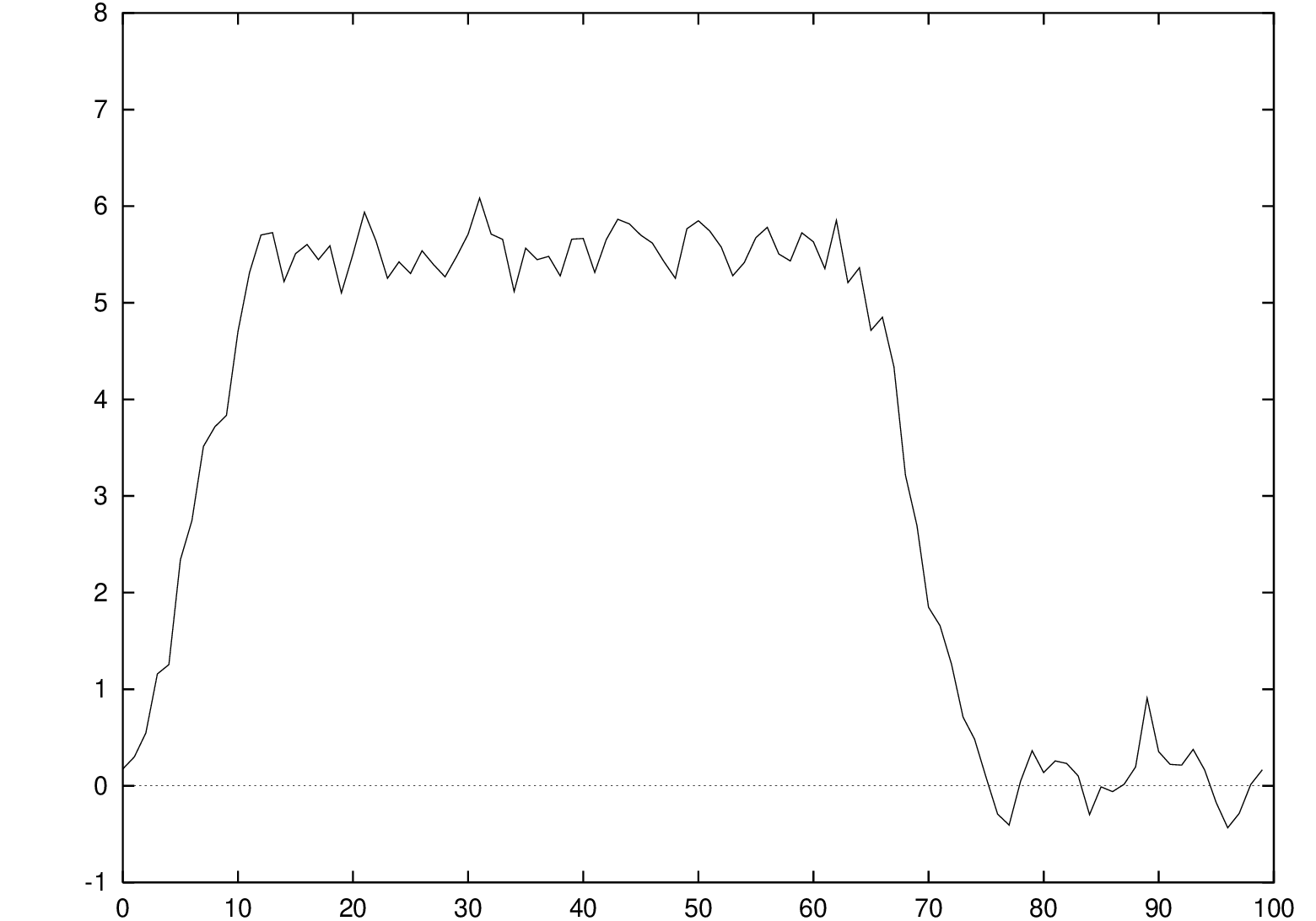,height=1.5in}
    }
  }
\hbox{\hspace{1.5in} ($t=1300$) \hspace{1.75in} ($t=1500$)} 
\vspace{9pt}
\caption{A series of snapshots of the lattice, $\phi(x)$ vs $x$, showing 
the growth of a fully formed bubble for a multiplicative  noise system.}
\label{figure7}
\end{figure}

Fig. \ref{figure6}
shows a series of frames which are snapshots of the lattice at different
times with
additive noise.  All results that follow pertain to solutions of the
dynamical equations rescaled to dimensionless form, as described in {\it 
  Appenxix B}.  Each point plotted is separated by $\triangle x=1$.
In the first two frames we see the field throughout the lattice 
fluctuating about the false vacuum at $\phi=0$, with preliminary indication of
the formation of a bubble in the region between 60 and 70 in the first
two frames.
The third frame shows an established, growing bubble. The fourth frame shows
a completely formed bubble.  Note that the small upward fluctuations in the
first frame in the regions around 5 and 95 are gone in the second frame.
 At the 
top of the profile the field fluctuates about the true vacuum value of
$\phi=5.0$.  
The expected ``$tanh$'' profile is discernible.\footnote{A numerical study of
the rate of expansion of the growing bubble can be found in Ref. \cite{haas}.} 
The matching of the profile at the 
boundaries is due to periodic boundary conditions. 
In Fig. \ref{figure7} we display the growth properties of bubbles using 
an example with multiplicative noise. When a fluctuation grows to occupy five 
contiguous lattice sites, we record that time as the nucleation time. 
\section{Results}
\subsection{Identifying Nucleation Times}
\begin{figure}[p]
  \vspace{9pt}

  \centerline{\hbox{ \hspace{0.0in} 
    \psfig{file=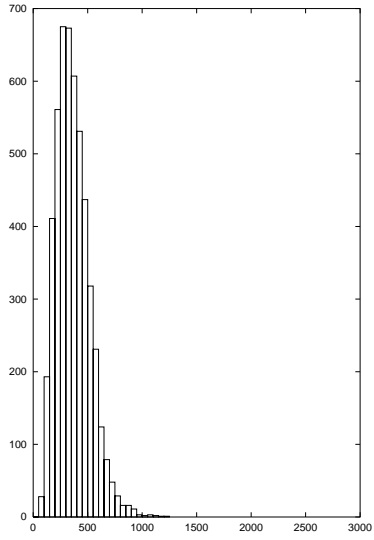,width=2in}
    \hspace{0.2in}
    \psfig{file=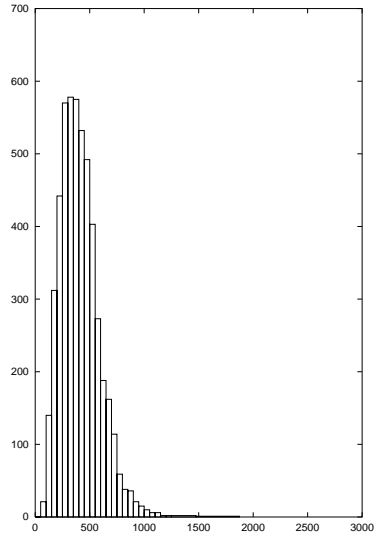,width=2in}
    }
  }
\vspace{9pt}
\hbox{\hspace{2.5in} (a) \hspace{2in}(b) } 
  \vspace{9pt}

  \centerline{\hbox{ \hspace{0.0in} 
    \psfig{file=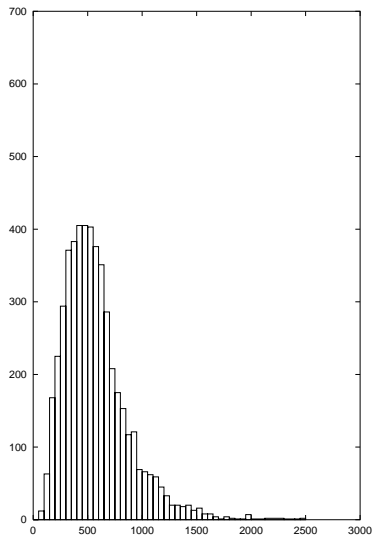,width=2in}
    \hspace{0.2in}
    \psfig{file=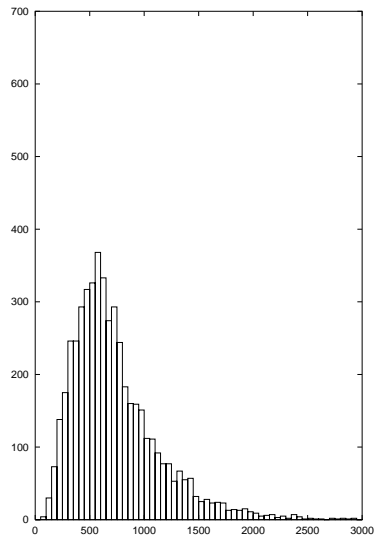,width=2in}
    }
  }
\hbox{\hspace{2.5in} (c) \hspace{2in} (d)} 
\vspace{9pt}
\caption{Distribution of nucleation times of 5000 bubbles at successively
lower temperatures, from the highest at (a) to the lowest at (d).} 
\label{figure8}
\end{figure}

\begin{figure}[htbp]
{\center \psfig{file=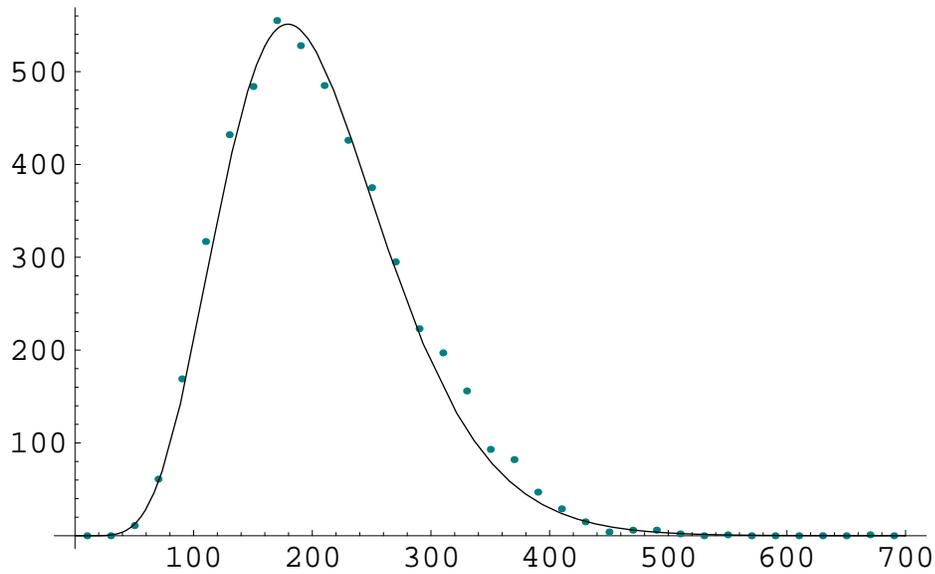,height=3in}}
\caption{Fit to Eqn. \ref{fit_fun} for an additive noise case with $T=1.5$ and 
$\alpha = 0.8$. From the fit we obtained  $a=6.8$ and  $\tau=26.2$}
\label{figure9}
\end{figure}

\begin{figure}[htbp]
{\center \psfig{file=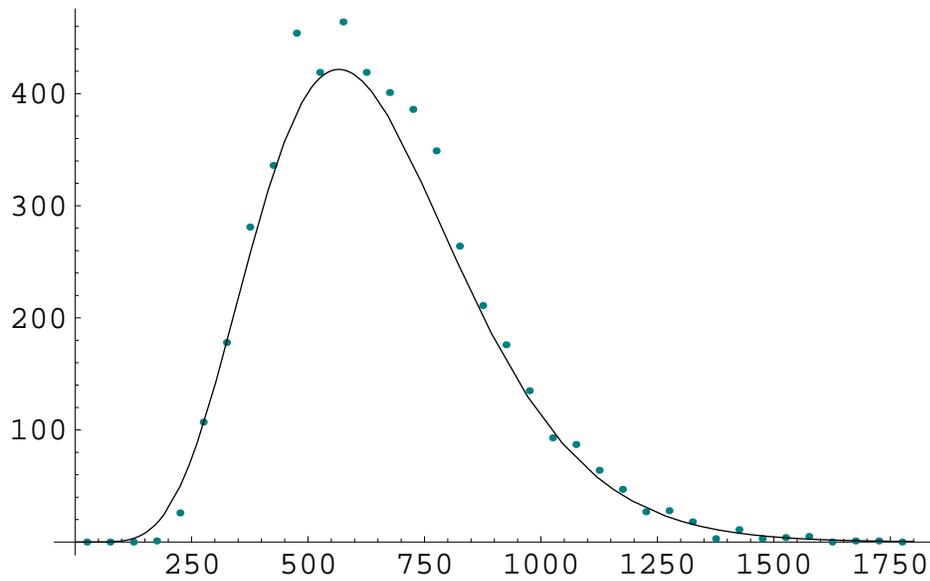,height=3in}}
\caption{Fit to Eqn. \ref{fit_fun} for a multiplicative noise case with $T=1.5$ and 
$\alpha = 0.8$. From the fit we obtained  $a=6.7$ and  $\tau=84.5$}
\label{figure10}
\end{figure}

\begin{figure}[htbp]
{\center \psfig{file=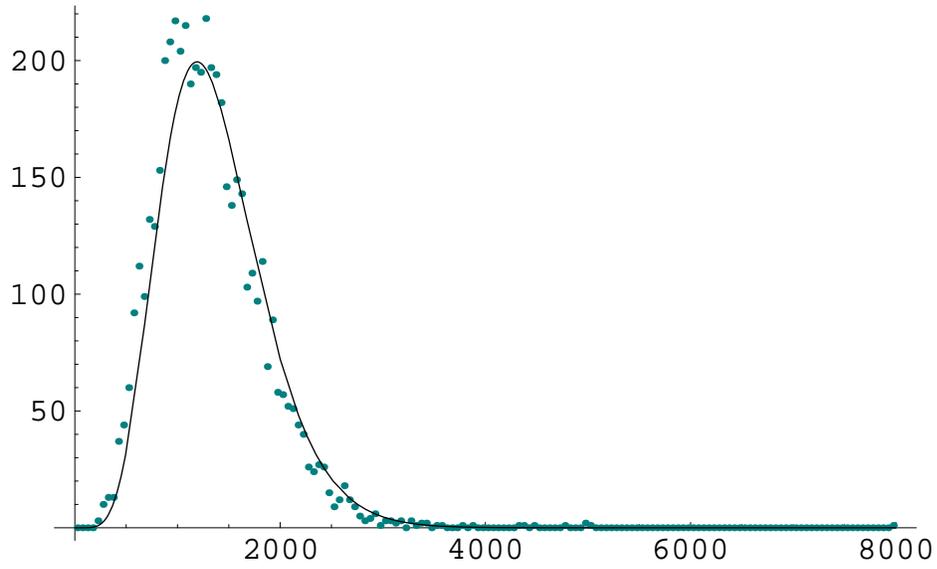,height=3in}}
\caption{Fit to Eqn. \ref{fit_fun} for an additive noise case with $T=1.2$ and 
$\alpha = 0.70$.}
\label{figure11}
\end{figure}

\begin{figure}[htbp]
{\center \psfig{file=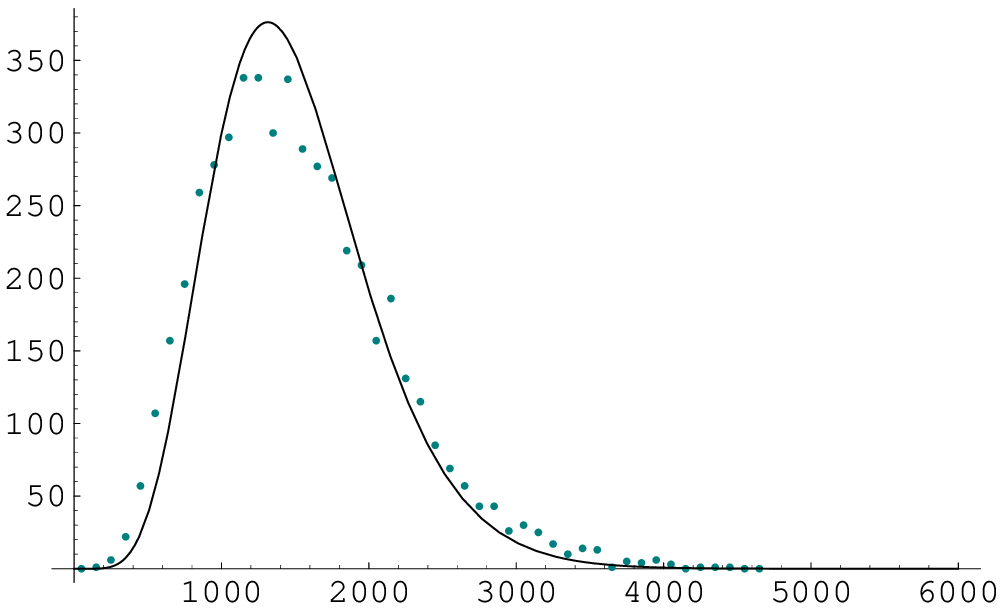,height=3in}}

\caption{Fit to Eqn. \ref{fit_fun} for a multiplicative noise case with $T=1.3$ and 
$\alpha = 0.74$.}
\label{figure12}
\end{figure}

The distribution of nucleation times for 5000 bubbles is 
shown in Fig. \ref{figure8}  at successively lower temperatures.
We see that the distribution is not symmetric; it rises quickly and there
is a long tail, which is especially pronounced at low temperatures. From
these distributions one has to extract nucleation times. 
We propose that the asymmetric distribution of times should be used as a whole.
Namely, we seek an appropriate function 
to fit the entire distribution of data at each temperature and
asymmetry parameter choice.\footnote{Based on remarks 
in the literature \cite{hume}\cite{algl} one might expect that the distribution
of nucleation times would be exponential.  Our results do not show this 
behavior.  Unfortunately, the details of the initial field configuration and
the actual distributions of nucleation times are not shown in these
references.  In the only paper
we know of that does show a nucleation time distribution, for a limited 
statistics sample in a $2+1$ dimensional model, the shape is in 
qualitative agreement with ours \cite{szep}.  These authors report exponential
fits to the {\it tails} of their distributions.  This procedure does not 
give the correct value for the nucleation time because the region of the 
distribution that is fit is not at long enough times to be in the true
exponential tail.  We develop this point further in the work below.}
We observe that the distributions rises to their maxima relatively quickly
compared to the time over which the tail of the distribution persists. 
We guess that a power law function of time will fit this part of the
distribution.  At long times, a damping function that ``overpowers a power
law'' is the falling exponential.  The simplest function that 
 has both of the above properties and has, at least naively, an appropriate
interpretation in our problem, is the classic ``waiting time'' form \cite{parzen}
%%%%%%%%%%%%%%%%%%%%%%%%%%%%%%%%%%%%%%%%%%%%%%%%%%%%%%%%%%%%%%%%%%%%%%%%%%%%%%
\begin{equation}
\frac{dN(t)}{dt}\delta t=K(\frac{t}{\tau})^{a}e^{-\frac{t}{\tau}}, 
\label{fit_fun}
\end{equation}
%%%%%%%%%%%%%%%%%%%%%%%%%%%%%%%%%%%%%%%%%%%%%%%%%%%%%%%%%%%%%%%%%%%%%%%%%%%%%
where $\delta t$ is the bin size of the histograms. 
The parameter $\tau$ is what we  use for the nucleation time. Both $a$ and
$\tau$ were obtained from fits to the distributions.  The normalization, $K$, can be
obtained from the requirement that the integral of Eq. \ref{fit_fun} gives the 
number of bubbles that were used to make the histogram.  Fig. \ref{figure9} shows
the least squares fit to this function using a typical additive noise case and 
Fig. \ref{figure10} shows the fit for a multiplicative noise case.  Figs. \ref{figure11} and \ref{figure12} show 
two more fits with different parameters.  These fits were made 
by setting $K=1$. From the normalization described above, we checked that this 
value is approximately correct in all cases. For the same
parameters in the models, and at the same temperature, the nucleation times
with multiplicative noise is always longer than that for the additive noise
case. 
 We emphasize that casting the equations of motion into dimensionless 
form, detailed in {\it Appendix B, } makes direct comparison between the models possible.  
%%%%%%%%%%%%%%%%%%%%%%%%%%%%%%%%%%%%%%%%%%%%%%%%%%%%%%%%%%%%%%%%%%%%%%%%%%%%
\subsection{Time vs $1/T$}
At finite temperature the nucleation rate per unit length, in the 
semi-classical approximation, is given by
%%%%%%%%%%%%%%%%%%%%%%%%%%%%%%%%%%%%%%%%%%%%%%%%%%%%%%%%%%%%%%%%%%%%%%%%%%%%%
\begin{equation}
\Gamma =A e^{\frac{-S_E}{T}}
\label{nuc_rate}
\end{equation}
%%%%%%%%%%%%%%%%%%%%%%%%%%%%%%%%%%%%%%%%%%%%%%%%%%%%%%%%%%%%%%%%%%%%%%%%%%%%%
where $S_E$ is the Euclidean action corresponding to the bounce solution of the
equation of motion and $T$ is the temperature. A simple manipulation of this equation 
gives  
%%%%%%%%%%%%%%%%%%%%%%%%%%%%%%%%%%%%%%%%%%%%%%%%%%%%%%%%%%%%%%%%%%%%%%%%%%%%%
\begin{equation}
ln\: \tau=ln(\Gamma L)^{-1} =-ln(AL) + \frac{S_E}{T},
\end{equation}
%%%%%%%%%%%%%%%%%%%%%%%%%%%%%%%%%%%%%%%%%%%%%%%%%%%%%%%%%%%%%%%%%%%%%%%%%%%%%
where $L$ is the length of the lattice. We extracted the nucleation times
from the fits as described above and looked at the effects of temperature
on nucleation times. To study the fluctuations in the parameters in the fits
we generated several
data sets for a single set of parameters. We found little variation in the 
parameters among runs. The tables in {\it Appendix C} show a set of  fit
parameters 
from different runs.
The fluctuations in the parameters $a$ and $\tau$ are small, being less that
two percent of the central value. The value of $a$ is especially stable from
one run to another.  We emphasize that, when normalized to 1, the average of
the fit function is $\Gamma[a+1]\times \tau$.  This should agree with the 
average of the data; {\it within (small) errors, it does in every case}.
                   
We also looked at the fluctuations in the fits at a lower
temperature and found that all the fluctuations in the parameters were within
five percent of the central value.  In fact most were within two percent, 
with $a$ showing smaller variation than $\tau$, as in the examples shown in the
tables in {\it Appendix C}.

\begin{figure}[htbp]
\center{\psfig{file=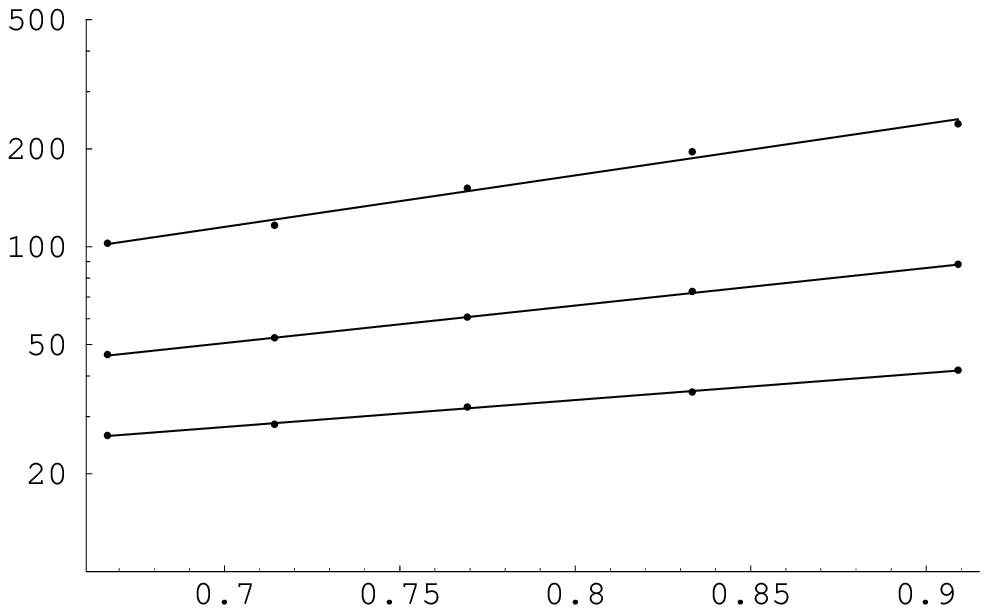,height=3in}}
\caption{A log-linear plot of $\tau$ vs. $T^{-1}$ for additive noise with different asymmetry 
parameters, $\alpha$=.70, .74 and .80 from the upper to the lower curve. The fits
are made with $5\%$ error bars. The confidence level of the fits are $85\%$, $98\%$ 
and $99\%$ respectively.}
\label{figure13}
\end{figure}

\begin{figure}[htbp]
\center{\psfig{file=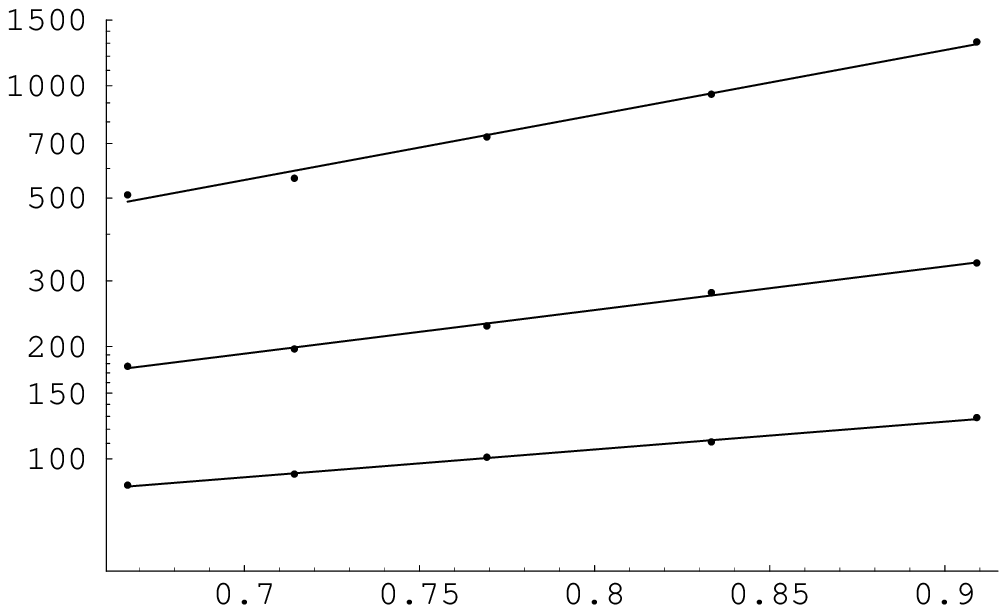,height=3in}}
\caption{A log-linear plot of $\tau$ vs. $T^{-1}$ for multiplicative noise with different asymmetry 
parameters, $\alpha$=.70, .74 and .80 from the upper to the lower curve.  The fits
are made with $5\%$ error bars. The confidence level of the fits are $95\%$, $98\%$ 
and $99\%$ respectively.}
\label{figure14}
\end{figure}

Using a global error estimate of five percent in the parameter
$\tau$ to cover all of the cases we studied, 
we fitted straight lines to the data plots of nucleation times vs inverse
temperature. We show the results for the 
additive model in Fig. \ref{figure13}.  The confidence levels of the fits are $85\%$, $98\%$ and $99\%$
for $\alpha=.70,\: .74$ and $.80$ respectively.  In Fig. \ref{figure14}. we show the corresponding result for the
multiplicative  noise model. The confidence levels of the fits are $95\%$, $98\%$ and $99\%$ 
for $\alpha=.70,\:.74$ and $.80$ respectively. 
Both plots show strong evidence that the relationship between $ln\:\tau$ and $1/T$
is linear, reproducing
the semiclassical description of the finite temperature quantum system, 
Eqn. \ref{nuc_rate}. 

We see that for both additive and multiplicative noise, the larger values of
the asymmetry parameter drive nucleation at a faster rate.
This is in agreement with nucleation theory:  
the action for the bounce, which appears in the exponential of the decay rate,
is inversely proportional to the energy difference between the true and false
vacua.  The plots also show that for a given asymmetry parameter the nucleation
time is shorter at higher temperatures as one would expect. As mentioned above,
the plots also indicate the linear relationship between $ln\:\tau$ and $1/T$.
{\it The new twist is that $\tau$ is extracted by our analysis from the full
distribution as a ``waiting time''}.  For long enough times, the waiting
time distribution, Eq. 16, has an exponential tail with decay constant $\tau$. 
But this is in the time regime $t\gg\tau a\times ln(t/\tau)$, far beyond the
point where we run out of simulated events.  It would take enormous statistics
to probe this regime.

\section{Discussion and Conclusion}
We have studied false vacuum decay in $1+1$ dimensions using Langevin equations on a lattice. We 
looked at cases where the stochastic term in the equation was in the form of 
 additive noise, previously studied in \cite{hume} in $1+1$, and in the form of
the theoretically motivated multiplicative noise case\cite{gleiserramos}.  The
cosmological applications of multiplicative noise have been discussed in\cite{habib}.
In past work \cite{hume}\cite{algl}\cite{szep}
decay times were fit by exponential distributions. We found distributions of 
nucleation times that were \emph{not exponential}, as described above. We
found that a
$\Gamma$, or waiting times, fit with effective multiple incidence of about 6,
gives a good description of the whole distribution of nucleation times for
every case studied.
In principle, the tail of the fitting function, Eq. 16, is an exponential
with decay time $\tau$.  In practice, the time must satisfy $t\gg \tau a\times
ln(t/\tau)$, which far beyond the end of our statistics, even with $5,000$
bubble samples.
At shorter times, well past the peak of the distribution but
where data is still sufficient in the falling tail, we can find reasonable,
approximate exponential fits to the data.   
However, these fits to the 
{\it exponential tails} do not give the same nucleation times as the
 $\Gamma$ distribution, in fact
the tails underestimate the nucleation times. Fitting the tails of the
 distribution also requires one to
discard most of the data and hence potential valuable information on the 
dynamics of nucleation. We found that 
for both the additive noise and multiplicative
noise Langevin models, the semiclassical relationship between nucleation time
and temperature is reproduced in 
the range of parameters studied. For the same set of parameters the 
multiplicative noise equations give nucleation times that are 
longer than in the additive case by factors of three or more. 
To see why this might be so, note that the location of the barrier of
the potential, and consequently its stable minimum, are at scaled field
values significantly greater than one. So
apparently the quadratic field dependence of the viscosity term in the
multiplicative case significantly inhibits the thermal hopping process, 
more than compensating for the enhancement from the linear field
dependence of the driving term. We tested this idea by varying the
value the field had to attain for a bubble to be declared formed.  We
expect that the fractional increase in nucleation time as the field
value increases should be greater in the multiplicative  noise case than 
in the additive.  We
looked at three different field value requirements, $\phi=5.0,\:4.0$,
and $3.0$.  Tables \ref{table0a} and \ref{table0b} show that for the
same increase 
in the field value criterion, the nucleation time increases by about $4\%$
in the additive noise model and $12\%$ in the multiplicative noise
model.  While this is not a proof of our conjecture, the trend 
demonstrates the importance of the field dependent viscosity in slowing
down the nucleation rate.

The $\Gamma$ distribution which fits the data is the waiting time distribution
for a fixed number$(>1)$ of
events that obey Poisson statistics. The fit parameter $a\simeq 6$ 
is fairly constant throughout the
range of input parameters studied, and this points to some quasi-universal 
feature in the models.  The observation that it is of the order of the
number of lattice sites, 5, that determines our bubble formation criterion,
which is consistent with the field correlation of the initial configuration, 
may be a clue to the interpretation of the 
insensitivity of the value of a to the choice of model and of the input
parameters for a given model.
The full interpretation of the successful application of the $\Gamma$ 
distribution to this problem and the significance of the approximate
universality of the parameter $a$ are currently under study. 
%%%%%%%%%%%%%%%%%%%%%%%%%%%%%%%%%%

\section{Appendix A \\ The fluctuation Dissipation Theorem}
The fluctuation-dissipation theorem\cite{kubo}\cite{chandrasekhar}, as the 
name implies, relates the 
stochastic fluctuations of a system to its dissipative or irreversible properties.  
As described earlier its application ranges from classical phenomena like Brownian motion to the far more
complicated case of a scalar field coupled to a thermal bath. For a general discussion see \cite{callenwelton}.
In numerical studies like the one reported here, the Langevin equation on its own does not
completely describe the coupling of the field to the thermal bath. The fluctuation-dissipation relation, Eqn. 
\ref{f_d}, is used to connect the stochastic and dissipation terms and hence ensure a realistic simulation
if the thermal system.  The delta function correlations are implemented numerically by sampling the
distribution of fluctuations at each lattice space site and at each time step in propagating the
solution of the equation of motion.  The exact form of the noise distribution and, in particular,  
its dependence on the viscosity parameter is important because from this we obtain the correct numerical 
value of temperature.  We describe here in some detail our implementation of the fluctuation
dissipation theorem and the noise distribution we sampled in the study.

The probability distribution of the noise is assumed to be Gaussian of the form,
%%%%%%%%%%%%%%%%%%%%%%%%%%%%%%%%%%%%%%%%%%%%%%%%%%%%%%%%%%%%%%%%%%%%%%%%
\begin{equation}
        dP[\xi]=\frac{1}{\sqrt{2\pi}\sqrt{2\eta T}}{\it D}\xi e^{-\frac{1}
{2}\int\frac{\xi^2 (x,t)}{2\eta T}dxdt}.
\end{equation}
%%%%%%%%%%%%%%%%%%%%%%%%%%%%%%%%%%%%%%%%%%%%%%%%%%%%%%%%%%%%
For convenience define $\theta=\sqrt{2\eta T}$ and $N=1/\sqrt{2\pi}\theta$. Then,
%%%%%%%%%%%%%%%%%%%%%%%%%%%%%%%%%%%%%%%%%%%%%%%%%%%%%%%%%%%%%%%%%%%%%%%%%%%%
\begin{equation}
<\xi(x,t)\xi(x',t')>=N\int D\xi \xi(x,t)\xi(x',t')e^{-\frac{1}{2}\int dx''dt''
\frac{\xi(x'',t'')^2}{\theta^2}}.
\end{equation}
%%%%%%%%%%%%%%%%%%%%%%%%%%%%%%%%%%%%%%%%%%%%%%%%%%%%%%%%%%%%%%%%%%%%%%%%%%
Introduce a source for the noise, $J(x,t)$, in the exponent.
%%%%%%%%%%%%%%%%%%%%%%%%%%%%%%%%%%%%%%%%%%%%%%%%%%%%%%%%%%%%%%%%%%%%%%%%%%
\begin{eqnarray}
<\xi(x,t)\xi(x',t')> &=&N\int D\xi\: \xi(x,t)\xi(x',t')\times\cr
& &e^{-\frac{1}{2}\int dx''dt''
\frac{\xi(x'',t'')^2}{\theta^2}\:+\: \int dx''dt'' J(x'',t'')\xi(x'',t'')}\\
&=&N\int D\xi \frac{\delta}{\delta J(x,t)}\frac{\delta}{\delta J(x',t')}\times\\
& &e^{-\frac{1}{2}\int dx''dt''
\frac{\xi(x'',t'')^2}{\theta^2}\:+\: \int dx''dt'' J(x'',t'')\xi(x'',t'')}
\end{eqnarray}

%%%%%%%%%%%%%%%%%%%%%%%%%%%%%%%%%%%%%%%%%%%%%%%%%%%%%%%%%%%%
Now re-define $\tilde\xi(x,t)$ to be $\xi(,x,t)=\xi(x,t)-\theta^2 J(x,t)$. 
%%%%%%%%%%%%%%%%%%%%%%%%%%%%%%%%%%%%%%%%%%%%%%%%%%%%%%%%%%%%
Then, \\
$-\frac{1}{2}\int dx''dt''\frac{\tilde\xi(x'',t'')^2}{\theta^2}$
\begin{equation}
%-\frac{1}{2}\int dx''dt''\frac{\tilde\xi(x'',t'')^2}{\theta^2}=
=-\frac{1}{2\theta^2}\int dx'' dt''(\xi(x,t)^2-
2\theta^2J(x'',t'')\xi(x'',t'')+\ \theta^4 J(x'',t'')^2).
\end{equation}
%%%%%%%%%%%%%%%%%%%%%%%%%%%%%%%%%%%%%%%%%%%%%%%%%%%%%%%%%%%%
Therefore,
%%%%%%%%%%%%%%%%%%%%%%%%%%%%%%%%%%%%%%%%%%%%%%%%%%%%%%%%%%%%
\begin{eqnarray}
<\xi(x,t)\xi(x',t')>&=&N\int D\tilde\xi \frac{\delta}{\delta J(x,t)}\frac{\delta}{\delta J(x',t')}
\times \\
&&e^{-\frac{1}{2}\int dx''dt''
\frac{\tilde\xi(x'',t'')^2}{\theta^2}\:+\: \frac{\theta^2}{2}\int dx''dt'' J(x'',t'')^2} \\
&=&N\sqrt{2\pi}\theta\frac{\delta}{\delta J(x,t)}\frac{\delta}{\delta J(x',t')}
e^{\frac{\theta^2}{2}\int dx''dt'' J(x'',t'')^2} \\
&=&N\sqrt{2\pi}\theta^3\frac{\delta}{\delta J(x,t)}
J(x',t')e^{\frac{\theta^2}{2}\int dx''dt'' J(x'',t'')^2}.
\end{eqnarray}

%%%%%%%%%%%%%%%%%%%%%%%%%%%%%%%%%%%%%
In the limit $J(x,t)\to 0$ we get, after taking the next functional derivative,
%%%%%%%%%%%%%%%%%%%%%%%%%%%%%%%%%%%%%
\begin{equation}
<\xi(x,t)\xi(x',t')>=N\sqrt{2\pi}\theta^3\delta(x-x')\delta(t-t')=2\eta T\delta(x-x')\delta(t-t')
\end{equation}
%%%%%%%%%%%%%%%%%%%%%%%%%%%%%%%%%%%%%
\section{Appendix B \\ Re-Scaling The equations of  Motion }

The parameters $\eta$ and $\xi$ have different units in the additive and
multiplicative noise equations of motion. To compare the results of 
our numerical study we rescale the equations of motion so that
all quantities appearing in them are dimensionless.  

Consider the potential, 
\begin{equation}
V(\phi)=\frac{m^2}{2}\phi^2-\frac{\alpha}{3}\phi^3+\frac{\lambda}{4}\phi^4.
\end{equation}
$V(\phi)$ appears in the Hamiltonian density, $\cal H$. 
\begin{equation}
{\cal H}=\frac{1}{2}\pi^2+\frac{1}{2}(\nabla\phi)^2+V(\phi)
\end{equation}
The Hamiltionian, $H$, which has the dimension of energy is obtained
from $\cal H$ as,
\begin{equation}
H=\int dx\cal\: H.
\end{equation}
$V(\phi)$ therefore has dimension of $EL^{-1}$.  From the term $(\nabla\phi)^2$ in $\cal H$ we get $[\phi]=(EL)^{\frac{1}{2}}$ where $[\cdot]$ means 
``dimension of''.  We also have the following;

\begin{eqnarray}
[m]&=& L^{-1}  \cr
[\lambda]&=& E^{-1}L^{-3}\cr 
[\alpha]&=& E^{-\frac{1}{2}}L^{-\frac{5}{2}}, \cr
[\eta_a]&=& L^{-1},\cr
[\xi_a]&=& E^{\frac{1}{2}}L^{-\frac{3}{2}}. \cr
\end{eqnarray}

The subscript $a$ refers to additive noise.  For the multiplicative 
noise case $\eta_m$ and $\xi_m$ have the dimensions,
\begin{eqnarray}
[\eta_m]&=&E^{-1}L^{-2},\cr
[\xi_m]&=&L^{-2}.
\end{eqnarray}

We now rescale the field and parameters appearing in the equation of motion.
For the purpose of discussion let $\phi_o$ be the value of the field in the true vacuum.\footnote{It can also
be chosen to be the value that the field takes at the top of the barrier of the 
potential or the value of the field at a zero of the potential.}  When the 
thin wall approximation is not applicable the field value at the escape point
at the right of the barrier is the more important quantity.  For the
additive noise case we have,
\begin{equation}
\frac{1}{\phi_0 m^2}\frac{\partial^2\phi}{\partial t^2}-
\frac{1}{\phi_0 m^2}\frac{\partial^2\phi}{\partial x^2}+
\frac{\eta_a}{\phi_0 m^2}\frac{\partial\phi}{\partial t}=
-\frac{\phi}{\phi_0}+\frac{\alpha\phi_o}{m^2}\frac{\phi^2}{\phi_o^2}
-\frac{\lambda\phi_0^2}{m^2}\frac{\phi^3}{\phi_0^3}+\frac{\xi_a}{\phi_0 m^2}.
        \label{}
\end{equation}

We now make the following definitions,

\begin{eqnarray}
\tilde\phi &=& \frac{\phi}{\phi_0},\cr
\tilde\eta_a &=& \frac{\eta_a}{m},\cr
\tilde\lambda &=& \frac{\lambda\phi_0^2}{m^2},\cr
\tilde\alpha &=& \frac{\alpha\phi_0}{m^2},\cr
\tilde\xi_a  &=& \frac{\xi_a}{\phi_0 m^2},\cr
\tilde x &=&m\:x,\cr
\tilde t &=& m\:t,\cr
\tilde T &=& \frac{T}{m\phi_0^2},
\end{eqnarray}

where $\tilde T$ follows from the fluctuation-dissipation relation,
Eqn. \ref{f_d}.  The rescaled equation of motion is,
\begin{equation}
\frac{\partial^2\tilde\phi}{\partial \tilde t^2}-
\frac{\partial^2\tilde\phi}{\partial \tilde x^2}+
\tilde\eta_a\frac{\partial\tilde\phi}{\partial \tilde t}=
-\tilde\phi+\tilde\alpha\tilde\phi^2
-\tilde\lambda\tilde\phi^3+\tilde\xi_a.
        \label{rescaled_add}
\end{equation}

We rescale the multiplicative noise case in a similar manner but the
dimensionless viscosity and noise parameters are,

\begin{equation}
\tilde\eta_m \:= \frac{\eta_m\phi_0^2}{ m}
\end{equation}
and 
\begin{equation}
\tilde\xi_m\:=\frac{\xi_m}{m^2}
\end{equation}

The dimensionless equation of motion is then
\begin{equation}
\frac{\partial^2\tilde\phi}{\partial \tilde t^2}-
\frac{\partial^2\tilde\phi}{\partial \tilde x^2}+
\tilde\eta_m\tilde\phi^2\frac{\partial\tilde\phi}{\partial \tilde t}=
-\tilde\phi+\tilde\alpha\tilde\phi^2
-\tilde\lambda\tilde\phi^3+\tilde\xi_m\tilde\phi.
        \label{rescaled_mul}
\end{equation}
In terms of the original parameters, the pre-scaled $T$ has the dimension 
of energy but it is dimensionless after re-scaling.  The key point is
that the parameters $m$, $\alpha$ and $\lambda$ in the scalar potential
can be used to rescale both equations to dimensionless forms.  We have
used $m$ and $\phi_0=(\alpha+\sqrt{\alpha^2-4m^2\lambda})/2$ to
illustrate the rescaling but other quantities of the same dimensions can
be used as noted.  All quantitive work is done with
Eqns. \ref{rescaled_add} and \ref{rescaled_mul}, so the values assigned
to $\tilde\alpha$, $\tilde\lambda$ determine the value of $\tilde\phi$
in the true vacuum in each case. With this understanding, the tilde
notation is not used in the text when referring to results from the
rescaled equations.  Now that the parameters
in both equations of motion are dimensionless, we can make direct comparisons of the two models.
%%%%%%%%%%%%%%%%%%%%%%%%%%%%%%%%%%%%%
\section{Appendix C\\Example Data and Fits Used to Determine Fluctuations in $\tau$ and $\lowercase{a}$}

\begin{figure}[htbp] 
{\center \psfig{file=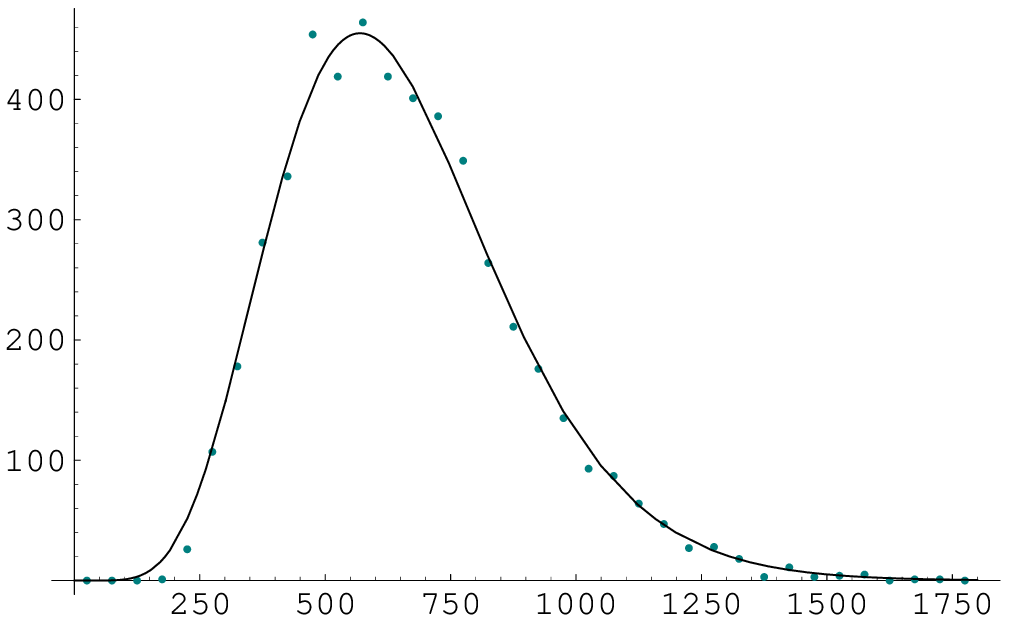,height=3in}}
\caption{Fit to Eqn. \ref{fit_fun} for a multiplicative noise case with $T=1.5$ and 
$\alpha = 0.8$}
\label{figure15}
\end{figure}

\begin{figure}[htbp]
{\center \psfig{file=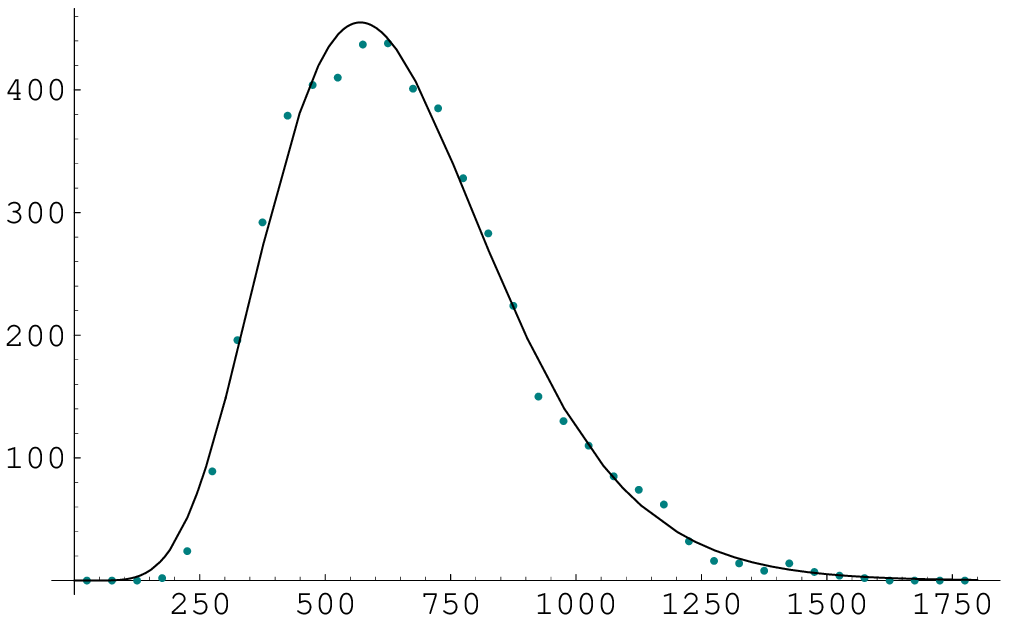,height=3in}}
\caption{Fit to Eqn. \ref{fit_fun} for a multiplicative noise case with $T=1.5$ and 
$\alpha = 0.8$}
\label{figure16}
\end{figure}

\begin{figure}[htbp]
{\center \psfig{file=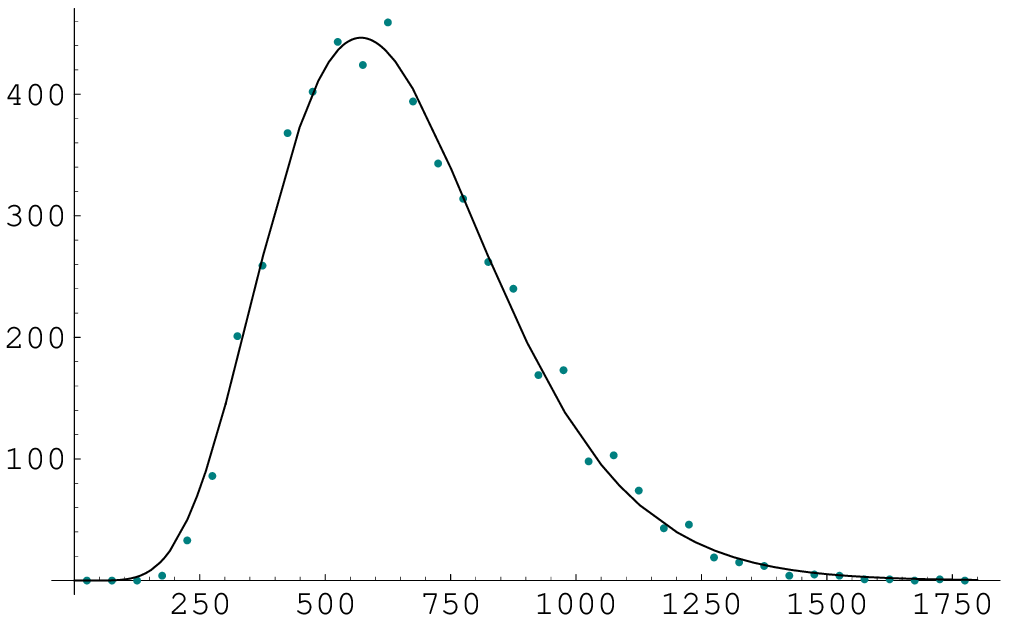,height=3in}}
\caption{Fit to Eqn. \ref{fit_fun} for a multiplicative noise case with $T=1.5$ and 
$\alpha = 0.8$.}
\label{figure17}
\end{figure}

\begin{figure}[htbp]
{\center \psfig{file=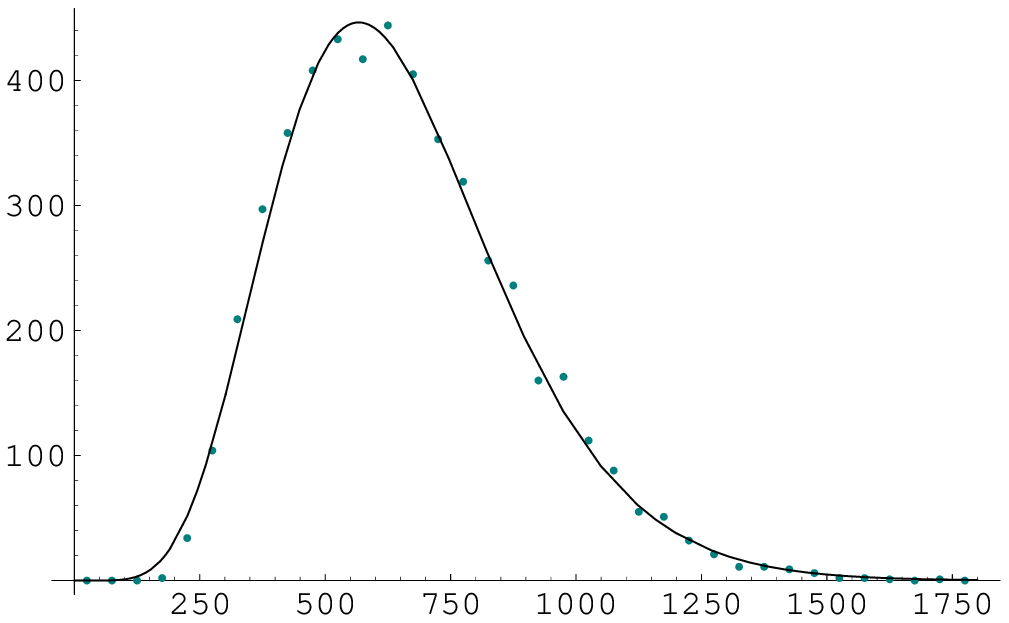,height=3in}}
\caption{Fit to Eqn. \ref{fit_fun} for a multiplicative noise case with $T=1.5$ and 
$\alpha = 0.8$.}
\label{figure18}
\end{figure}

\begin{figure}[htbp]
{\center \psfig{file=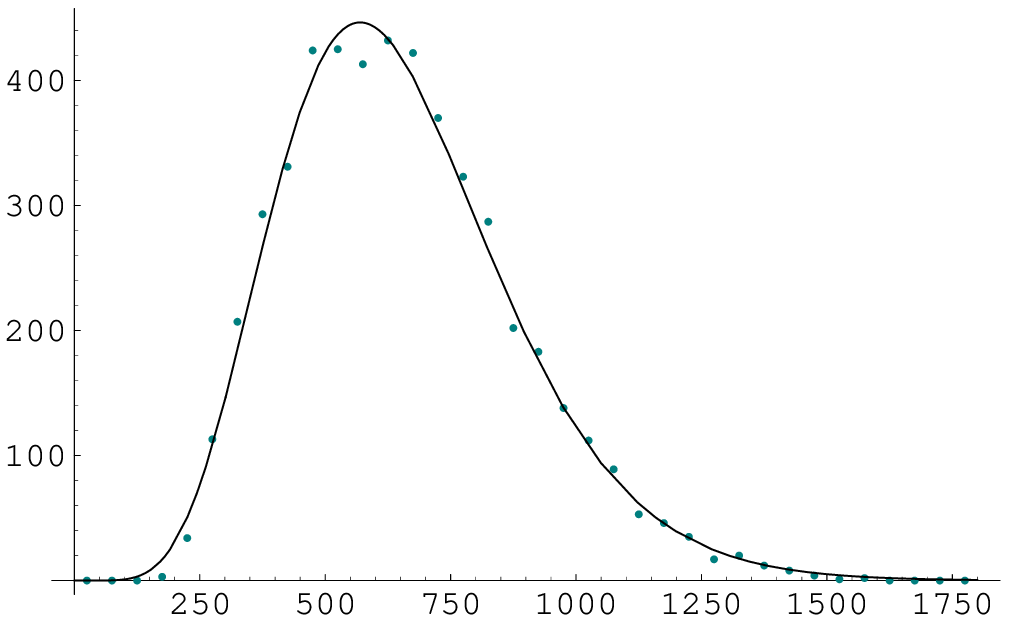,height=3in}}
\caption{Fit to Eqn. \ref{fit_fun} for a multiplicative noise case with $T=1.5$ and 
$\alpha = 0.8$.}
\label{figure19}
\end{figure}

\begin{figure}[htbp]
{\center \psfig{file=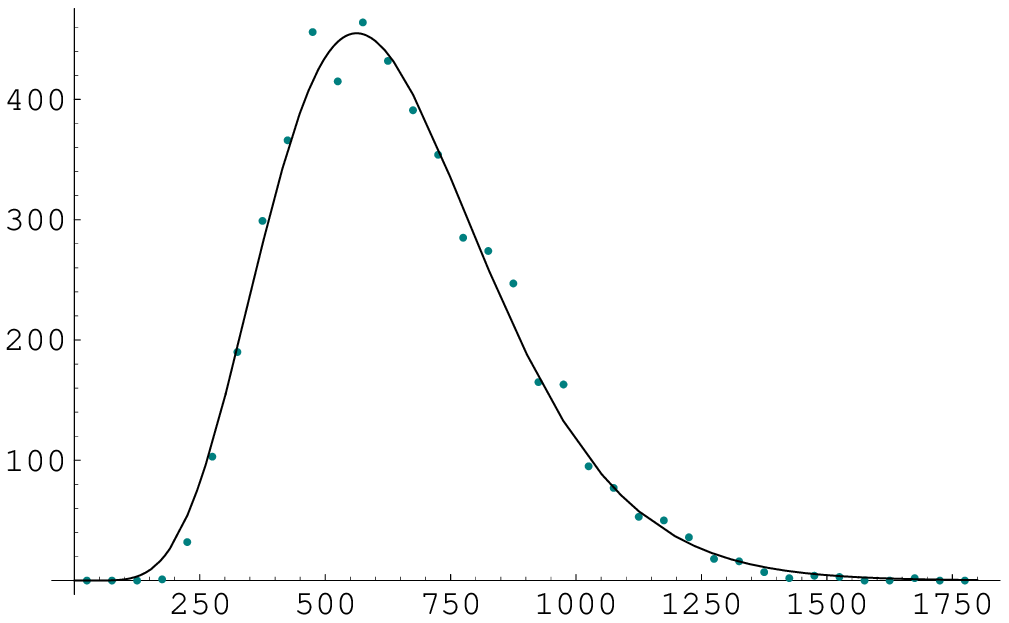,height=3in}}
\caption{Fit to Eqn. \ref{fit_fun} for a multiplicative noise case with $T=1.5$ and 
$\alpha = 0.8$.}
\label{figure20}
\end{figure}

Tables \ref{table1} through \ref{table6} and accompanying Figures \ref{figure15} through \ref{figure20} show examples of
data and fits that we used to study sample-to-sample fluctuations.  Error
estimates used to assign confidence levels to linear fits to the $ln\:(\tau)\:vs.\:T^{-1}$ plots, 
Figs 13 and 14, are based on these studies.

%%%%%%%%%%%%%%%%%%%%%%%%%%%%%%%%%%%%%%%%%%%%%%%%%%
\acknowledgements{
Conversations and suggestions from Ruslan Davidchack, Brian Laird, 
John Ralston, and  Krzysztof Kuczera at various stages of this work were 
of great value.  We thank Hume Feldman for numerous discussions on
his work in this field and for critical comments on some of our early
results. We thank Salman Habib for bringing to our attention his 
earlier work on multiplicative noise.  This work was supported
in part by U.S. Department of Energy grant number DE-FG03-98ER41079.
Facilities of the Kansas Institute for Theoretical and Computational
Science and especially the Kansas Center for Advanced Scientific Computing
were essential for this work.}
%%%%%%%%%%%%%%%%%%%%%%%%%%%%%%%%%%

%%%%%%%%%%%%%%%%%%%%%%%%%%%%%%%%%%%%%%%%%%%
%%%%%%%%%%%%%%%%%%%%%%%%%%%%%%%%%%%%%%%%%%%%%%%
%%%%%%%%%%%%%%%%%%%%%%%%%%%%%%%%%%%%%%%%%%%%%%%%%%%%%%%%%%%%%%%%%%%%%%%%%%%%%%
%%%%%%%%%%%%%%%%%%%%%%%%%%%%%%%%%%%%%%%%%%%%%%%
%%%%%%%%%%%%%%%%%%%%%%%%%%%%%%%%%%%%%%%%%%%%%%%%%%%%%%%%%%%%%%%%
%%%%%%%%%%%%%%%%%%%%%%%%%%%%%%%%%%%%%%%%%%%%%%%%%%%%%%%%%%%%%%%%%
%%%%%%%%%%%%%%%%%%%%%%%%%%%%%%%%%%%%%%%%%%%%%%%%%%
%%%%%%%%%%%%%%%%%%%%%%%%%%%%%%%%%%%%%%%%%%%%%%%%%%
%%%%%%%%%%%%%%%%%%%%%%%%%%%%%%%%%%%%%%%%%%%%%%%%%%%%%%%%%%%%%%%%%%%%%%%%%%%%%%
%%%%%%%%%%%%%%%%%%%%%%%%%%%%%%%%%%%%%%%%%%%%%%%%%%%%%%%%%%%%%%%%%%%%%%%%%%%
%%%%%%%%%%%%%%%%%%%%%%%%%%%%%%%%%%%%%%%%%%%%%%%%%%%%%%%%%%%%%%%%%%%%%%%%%%%

%%%%%%%%%%%%%%%%%%%%%%%%%%%%%%%%%%%%%%%%%%%%%%%%%%%%%%%%%%%%%%%%%%%%%%%%%%%
%%%%%%%%%%%%%%%%%%%%%%%%%%%%%%%%%%%%%%%%%%%%%%%%%%%%%%%%%%%%%%%%%%%%%%%%%%%%%
%%%%%%%%%%%%%%%%%%%%%%%%%%%%%%%%%%%%%%%%%%%%%%%%%%%%%%%%%%%%%%%%%%%%%%%%%%%%% 

%%%%%%%%%%%%%%%%%%%%%%%%%%%%%%%%%%%%%%%%%%%%%%%%%%%%%%%%%%%%%%%%%%%%%%%%%%
%%%%%%%%%%%%%%%%%%%%%%%%%%%%%%%%%%%%%%%%%%%%%%%%%%%%%%%%%%%%%%%%%%%%%%%%%%%%%%%%
%%%%%%%%%%%%%%%%%%%%%%%%%%%%%%%%%%%%%%%%%%%%%%%%%%%%%%%%%%%%%%%%%%%%%%%%% 
%%%%%%%%%%%%%%%%%%%%%%%%%%%%%%%%%%%%%%%%%%%%%%%%%%%%%%%%%%%%%%%%%%%%%%%%%
%%%%%%%%%%%%%%%%%%%%%%%%%%%%%%%%%%%%%%%%%%%%%%%%%%%%%%%%%%%%%%%%%%%%%%  
%%%%%%%%%%%%%%%%%%%%%%%%%%%%%%%%%%%%%%%%%%%%%%%%%%%%%%%%%%%%%%%%%%%%%%  
%%%%%%%%%%%%%%%%%%%%%%%%%%%%%%%%%%%%
\begin{table}[t]
%$\alpha=.80$
$$ \vbox{\vskip4pt
\halign{#\hfil\qquad&#\hfil\qquad&#\hfil\qquad&#\hfil\qquad \cr
\hline
$\phi$& $5.0$& $4.0$ & $3.0$\cr
\hline
$\tau$  &27.02$\pm$.18 &25.9$\pm$.2 & 24.8$\pm$.2\cr
$a$ &6.63$\pm$.005 &6.839$\pm$.004 & 6.545$\pm$.004\cr
$\tau_{avg}$&212$\pm$1 &204$\pm$1 & 196$\pm$1\cr
\hline
}}
$$

\caption{Parameters from fits for additive noise with $T=1.5$ and
  $\alpha=.80$, and $\lambda=.1$. Each column represents a different
  $\phi$ value which the field was required to reach before the
  nucleation time was recorded.}
\label{table0a}
\end{table}
%%%%%%%%%%%%%%%%%%%%%%%%%%%%%%%%%%%%
\begin{table}[t]
%$\alpha=.80$
$$ \vbox{\vskip4pt
\halign{#\hfil\qquad&#\hfil\qquad&#\hfil\qquad&#\hfil\qquad \cr
\hline
$\phi$& $5.0$& $4.0$ & $3.0$\cr
\hline
$\tau$  &84.24$\pm$.43 &73.94$\pm$.9 & 63.8$\pm$.6\cr
$a$ &6.74$\pm$.005 &6.796$\pm$.005 & 6.796$\pm$.004\cr
$\tau_{avg}$&654$\pm$3 &581$\pm$3 & 510$\pm$3\cr
\hline
}}
$$

\caption{Parameters from fits for multiplicative noise with $T=1.5$ and
  $\alpha=.80$, and $\lambda=.1$. Each column represents a different
  $\phi$ value which the field was required to reach before the
  nucleation time was recorded.}
\label{table0b}
\end{table}
%%%%%%%%%%%%%%%%%%%%%%%%%%%%%%%%%%%%
\begin{table}[t]
%$\alpha=.80$
$$ \vbox{\vskip4pt
\halign{#\hfil\qquad&#\hfil\qquad&#\hfil\qquad \cr

$a$ &	$\tau$&	$\tau_{avg}$\cr
6.83&	26.90&	213\cr
6.84&	26.71&	211\cr
6.84&	27.03&	212\cr
6.83&	26.94&	212\cr
6.84&	26.99&	213\cr
6.83&	27.22&	213\cr
6.83&	27.11&	213\cr
6.83&	27.35&	213\cr
6.83&	26.94&	212\cr
&$average$  & \cr
6.83&	27.02&	212\cr
&$\sigma$  & \cr
.005&	0.18&	1\cr}}
$$

\caption{Parameters from fits for additive noise with $T=1.5$ and $\alpha=.80$}
\label{table1}
\end{table}
%%%%%%%%%%%%%%%%%%%%%%%%%%%%%%%%%%%%%%%%%%%%%%%%%%%%%%%%%%%%%%%%%%%%%%%%%%%
\begin{table}
$$ \vbox{\vskip4pt
\halign{#\hfil\qquad&#\hfil\qquad&#\hfil\qquad \cr
$a$ &	$\tau$&	$\tau_{avg}$\cr
6.51	&48.27&	366\cr
6.52	&45.04&	359\cr
6.51	&44.62&	362\cr
6.51	&44.66&	362\cr
6.51	&44.61&	363\cr
6.51	&44.64&	362\cr
6.51	&44.56&	363\cr
6.51	&44.65&	362\cr
6.51	&44.66&	362\cr
6.51	&44.62&	363\cr
& $average$ & \cr
6.51	&45.03&	362\cr
&$\sigma$  & \cr
0.003	&1.15&	2\cr}}
$$

\caption{Parameters from fits for additive noise with $T=1.5$ and $\alpha=.74$}
\end{table}
%%%%%%%%%%%%%%%%%%%%%%%%%%%%%%%%%%%%%%%%%%%%%%%%%%%%%%%%%%%%%%%%%%%%%%%%%%%
\begin{table}[htbp]
%$\alpha=.70$
$$ \vbox{\vskip4pt
\halign{#\hfil\qquad&#\hfil\qquad&#\hfil\qquad \cr
$a$ &	$\tau$&	$\tau_{avg}$\cr
6.04	&103.62	&797\cr
6.05	&106.59	&787\cr
6.04	&106.59	&784\cr
6.05	&103.16	&791\cr
6.05	&105.10	&776\cr
6.05	&104.70	&786\cr
6.05	&102.77	&768\cr
6.04	&104.85	&786\cr

&$average$ & \cr
6.05	&104.67	&784\cr

& $\sigma$  & \cr
0.005	&1.44	&9\cr}}
$$

\caption{Parameters from fits for additive noise with $T=1.5$ and $\alpha=.70$}
\end{table}
%%%%%%%%%%%%%%%%%%%%%%%%%%%%%%%%%%%%%%%%%%%%%%%%%%%%%%%%%%%%%%%%%%%%%%%%%%%
\begin{table}[htbp]
$$ \vbox{\vskip4pt
\halign{#\hfil\qquad&#\hfil\qquad&#\hfil\qquad \cr

$a$ &	$\tau$&	$\tau_{avg}$\cr
6.74	&84.48	&654\cr
6.74	&83.91	&651\cr
6.74	&84.58	&657\cr
6.73	&84.98	&660\cr
6.73	&84.23	&654\cr
6.73	&84.64	&654\cr
6.74	&83.48	&650\cr
6.74	&84.13	&654\cr
6.74	&84.02	&652\cr
6.74	&83.99	&654\cr

&$average$ & \cr
6.74 &84.24	&654\cr

&$\sigma$  & \cr
0.005 &0.43&	3\cr}}
$$

\caption{Parameters from fits for multiplicative noise with $T=1.5$ and $\alpha=.80$}
\end{table}
%%%%%%%%%%%%%%%%%%%%%%%%%%%%%%%%%%%%%%%%%%%%%%%%%%%%%%%%%%%%%%%%%%%%%%%%%%%
\begin{table}[htbp]
$$ \vbox{\vskip4pt
\halign{#\hfil\qquad&#\hfil\qquad&#\hfil\qquad \cr

$a$ &	$\tau$&	$\tau_{avg}$\cr
6.31	&171.50	&1303\cr
6.31	&166.91	&1313\cr
6.30	&173.00	&1330\cr
6.32	&168.67	&1300\cr
6.31	&173.78	&1309\cr
6.30	&171.67	&1310\cr
6.31	&173.16	&1305\cr
6.31	&168.47	&1306\cr
6.31	&171.14	&1316\cr
6.31	&171.72	&1323\cr
&$average$&  \cr
6.309	&171.002&	1311\cr
&$\sigma$  & \cr
0.006	&2.27	&9
\cr}}
$$

\caption{Parameters from fits for multiplicative noise with $T=1.5$ and $\alpha=.74$}
\end{table}
%%%%%%%%%%%%%%%%%%%%%%%%%%%%%%%%%%%%%%%%%%%%%%%%%%%%%%%%%%%%%%%%%%%%%%%%%%%
\begin{table}[htbp]
$$ \vbox{\vskip4pt
\halign{#\hfil\qquad&#\hfil\qquad&#\hfil\qquad \cr
$a$ &	$\tau$&	$\tau_{avg}$\cr
6.08	&504.00	&3485\cr
6.06	&509.00	&3754\cr
6.09	&500.79	&3451\cr
6.07	&511.79	&3513\cr
6.09	&495.37	&3495\cr
6.09	&490.05	&3493\cr
6.08	&496.60	&3487\cr
6.07	&505.51	&3510\cr
6.07	&513.70	&3508\cr
6.07	&520.18	&3542\cr
&$average$&  \cr
6.08	&504.70	&3524\cr
&$\sigma$  & \cr
0.011	&9.27	&84\cr
\cr}}
$$

\caption{Parameters from fits for multiplicative noise with $T=1.5$ and $\alpha=.70$}
\label{table6}
\end{table}
%%%%%%%%%%%%%%%%%%%%%%%%%%%%%%%%%%%%%%%%%%%%%%%%%%%%%%%%%%%%%%%%%%%%%%%%%%%

\end{document}